\documentclass[11pt]{article}
\usepackage{theorem}
\usepackage{latexsym}
\usepackage{amssymb}
\usepackage{amsmath,mathpazo,color}
\newtheorem{thm}{Theorem}[section]

\newtheorem{prop}[thm]{Proposition}
\newtheorem{lem}[thm]{Lemma}

\theorembodyfont{\rm}
\newtheorem{defn}[thm]{Definition}
\newtheorem{ex}[thm]{Example}

\newtheorem{rem}[thm]{Remark}

\def\dint  {{\displaystyle\int}}
\def\fin   {\hfill{$\Box$}\vspace{5mm}}
\def\proof {\noindent{\it Proof.}$\quad$$\quad$}
\def\l     {\left}
\def\r     {\right}
\def\<     {\langle}
\def\>     {\rangle}
\def\as    {\mathop{\mathrm{a.s.}}\nolimits}
\def\co    {\mathop{\mathrm{co}}\nolimits}

\def\d     {\mathrm{d}}
\def\wt#1  {\widetilde#1 }
\def\wh#1  {\widehat#1 }
\def\ol#1  {\overline#1 }
\def\ul#1  {\underline#1 }
\def\calA  {{\cal A}}

\def\calF  {{\cal F}}

\def\calP  {{\cal P}}
\def\calQ  {{\cal Q}}
\def\calR  {{\cal R}}
\def\calS  {{\cal S}}

\def\bbE   {\mathbb E}
\def\bbF   {\mathbb F}

\def\bbN   {\mathbb N}
\def\bbP   {\mathbb P}
\def\bbR   {\mathbb R}
\def\bbZ   {\mathbb Z}
\def\ve    {\varepsilon}
\def\vt    {\vartheta}
\def\olm   {\overline{m}}
\def\olL   {\overline{L}}
\def\olM   {\overline{M}}

\begin{document}
\title{Good deal bounds with convex constraints}
\author{Takuji Arai\footnote{
        Department of Economics, Keio University,
        Minato-ku, Tokyo, 108-8345 Japan}}
\maketitle

\begin{abstract}
We investigate the structure of good deal bounds,
which are subintervals of a no-arbitrage pricing bound,
for financial market models with convex constraints
as an extension of Arai and Fukasawa \cite{AF}.
The upper and lower bounds of a good deal bound are naturally described by
a convex risk measure.
We call such a risk measure a good deal valuation; and study its properties.
We also discuss superhedging cost and Fundamental Theorem of Asset Pricing
for convex constrained markets.

\end{abstract}
       
\setcounter{equation}{0}
\section{Introduction}
For a given contingent claim in an incomplete financial market,
its price is not determined uniquely under the no-arbitrage framework.
Only a pricing bound, called a no-arbitrage pricing bound,
is provided.
For a concrete explanation,
letting $L$ be a linear space of measurable functions representing all future
cash-flows, we describe our market with $M\subset L$
the set of $0$-attainable claims,
that is, future payoffs which investors can replicate completely
with $0$ initial cost.
In this paper, $M$ is assumed to be convex, but not necessarily a cone.
Defining a functional on $L$ as
\[
\rho^0(x):=\inf\{r\in\bbR|\mbox{ there exists $m\in M$ such that }r+m+x\geq0\}
\]
for $x\in L$, we obtain that the superhedging cost and
the no-arbitrage pricing bound for claim $x$ are given by $\rho^0(-x)$ and
$[-\rho^0(x), \rho^0(-x)]$ respectively.

Generally speaking, the no-arbitrage pricing bound is too wide to be useful
as the collection of candidate prices from a practical point of view.
Thus, we focus on narrowing the interval of candidate prices.
That is, excluding prices with low risk, called good deals,
from the no-arbitrage pricing bound,
we construct a sharper pricing bound, called a good deal bound.
Whether a price is a good deal, depends on the investor's risk preference.
For example, Cochrane and Sa\'a-Requejo \cite{CS00} defined a good deal bound
by excluding prices with high Sharpe ratio.
We do not treat any specific good deal bound in this paper.
Instead, we are interested in the entire structure of good deal bounds.
We regard the upper and lower bounds of a good deal bound
as functionals on $L$, denoted by $a$ and $b$ respectively.
Thus, the interval $[b(x), a(x)]$,
which is a subinterval of $[-\rho^0(x), \rho^0(-x)]$,
formulates a good deal bound.
As in Arai and Fukasawa \cite{AF}, defining $\rho_a(-x):=a(x)$,
it is natural that $\rho_a$ becomes a convex risk measure.
We call it a good deal valuation (GDV, hereafter).
That is, a GDV is defined as a convex risk measure describing
the upper bound of a good deal bound.
Similarly, defining $-\rho_b(x):=b(x)$, we can see that $\rho_b$ is also a GDV,
that is, it can describe the upper bound of a different good deal bound.

Our aim of this paper is to study GDVs for the case where $M$ is convex
along with the argument of \cite{AF},
in which they studied GDVs thoroughly when $M$ is a convex cone.
They showed that $\rho^0$ is a coherent risk measure; and enumerated
equivalent conditions for the existence of a GDV.
Besides, for a given convex risk measure, they gave a set of
equivalent conditions to be a GDV.
In particular, they showed that $\rho$ is a GDV if and only if
it is a risk indifference price.
Furthermore, they extended the Kreps-Yan-type
Fundamental Theorem of Asset Pricing (FTAP, hereafter) to
the equivalence between the no-free-lunch condition (NFL, hereafter) and
the existence of a relevant GDV.
In addition to \cite{AF}, there is much literature on good deal bounds
from the point of view of risk measures, say,
Bion-Nadal \cite{BN09}, Bion-Nadal and
Di Nunno \cite{BNDN}, Jaschke and K\"uchler \cite{JK01} and Staum \cite{S04}.
But, no one studied good deal bounds for markets with convex constraints,
whereas such models appear frequently in mathematical finance, say,
illiquid market models, models with borrowing constraints and so on.
Indeed, there is much literature treating models with convex constraints:
Cuoco \cite{Cuoco}, Cvitani\'c and Karatzas \cite{CK92} and \cite{CK93},
Karatzas and Kou \cite{KK96}, Larsen and \v{Z}itkovi\'{c} \cite{LZ13},
Pennanen \cite{P1} and \cite{P2}, Pennanen and Penner \cite{PP}, and so forth.
See also examples introduced in Subsection 2.1.

Our main contribution is threefold as follows:
\begin{enumerate}
\item 
We begin with a study for the functional $\rho^0$, since it expresses
the upper and lower bounds of the no-arbitrage pricing bound.
As seen in \cite{AF}, $\rho^0$ is given as a coherent risk measure
when $M$ is a convex cone.
In this case, the set $\calQ_0$ of all probability measures $Q$ such that
$\sup_{m\in M}\bbE_Q[m]=0$, plays a central role to discuss
not only $\rho^0$ but also GDVs.
On the other hand, excluding the cone property from $M$,
$\rho^0$ is no longer coherent in general.
In this setting, we need to consider, instead of $\calQ_0$,
the set, denoted by $\calQ$, of all probability measures $Q$ such that
$\sup_{m\in M}\bbE_Q[m]$ is finite.
In particular, we investigate properties of the largest minorant of $\rho^0$
with the Fatou property, since it is the first candidate of GDVs.

\item 
We shall enumerate equivalent conditions for the existence of a GDV;
and introduce a set of equivalent conditions for a given convex risk measure
to be a GDV.
In addition, we introduce an example of a GDV which is not
a risk indifference price.
This shows that the structure of good deal bounds is much different from
that for the case where $M$ is a convex cone.
Moreover, we give conditions for a GDV to be a risk indifference price;
and for a risk indifference price to be a GDV.

\item 
We deal with the Kreps-Yan-type FTAP.
Kreps \cite{K81} proved that, if $M$ is a convex cone, $\calQ_0\neq\emptyset$
is equivalent to the NFL, that is,
the weak closure of $M$ does not include any nonzero nonnegative claims.
Moreover, \cite{AF} showed that the existence of a relevant GDV
is equivalent to the NFL.
Thus, we expect naturally that, when $M$ is convex, the equivalence holds true
among the NFL, the existence of a relevant GDV and a condition related to
$\calQ$.
Indeed, we shall see the equivalence between the first two conditions, but
illustrate counterexamples for the last one.
Some variants of FTAP for constrained models have been introduced by
Carassus, Pham and Touzi \cite{CPT01},
Evstigneev, Sch\"urger and Taksar \cite{EST04},
\cite{P1}, Rokhlin \cite{R05}, Roux \cite{R11} and so on.
Thus, our contribution is to treat FTAP comprehensively
for models with convex constraints.
\end{enumerate}

An outline of this paper is as follows.
In Section 2, we describe our model; and prepare some terminologies and
mathematical preliminaries.
In particular, since we take an Orlicz space (or heart) as $L$,
we introduce some terminologies on Orlicz space.
We study $\rho^0$ in Section 3.
Section 4 is devoted to study properties of GDVs.
FTAP will be discussed in Section 5; and conclusions are given in Section 6.

\setcounter{equation}{0}
\section{Preliminaries}
Throughout this paper, we fix a complete probability space
$(\Omega, \calF, \bbP)$.
Note that we denote by $\bbN$ the set of all positive integers; and
by $L^0$ the set of all $\bbR$-valued measurable functions
on $(\Omega, \calF)$.
Moreover, for a set of measurable functions $X$,
we denote $X_+$ (resp. $X_-$)$:=\{x\in X|x\geq0\as$ (resp. $\leq$)$\}$.

We start with definitions of Young function, Orlicz heart and Orlicz space.

\begin{defn}
\begin{enumerate}
\item An even lower semi-continuous convex function
      $\Phi: \bbR\to\bbR\cup\{\infty\}$ is called a Young function,
      if it satisfies the following:
      \begin{enumerate}
      \item $\Phi(0)=0$,
      \item $\Phi(\alpha)\uparrow\infty$ as $\alpha\uparrow\infty$,
      \item $\Phi(\alpha)<\infty$ for $\alpha$ in a neighborhood  of $0$.
      \end{enumerate}
\item For a Young function $\Phi$, a space $M^\Phi$ of measurable functions
      on $(\Omega, \calF)$ defined as
      \[
      M^\Phi:=\{x\in L^0|\bbE[\Phi(cx)]<\infty\mbox{ for any }c>0\}
      \]
      is called Orlicz heart with $\Phi$.
      In addition, a space $L^\Phi$ defined as
      \[
      L^\Phi:=\{x\in L^0|\bbE[\Phi(cx)]<\infty\mbox{ for some }c>0\}
      \]
      is called Orlicz space with $\Phi$.
\item The complimentary function of $\Phi$ is defined as
      \begin{equation*}
      \Psi(\beta):=\sup_{\alpha\in\bbR}\{\alpha\beta-\Phi(\alpha)\}
      \end{equation*}
      for any $\beta\in\bbR$.
      Note that $\Psi$ is also a Young function.
\end{enumerate}
\end{defn}

\noindent
Any Young function is continuous on $[0,\infty)$ except for possibly
a single point at which it jumps to $+\infty$.
Both $M^\Phi$ and $L^\Phi$ are Banach lattices with norm
$\|x\|:=\inf\{c>0|\bbE[\Phi(x/c)]\leq1\}$
and pointwise ordering in the almost sure sense.
When $\Phi$ is finite, $L^\Phi=M^\Phi$ if and only if
we can find $c>0$ and $\alpha_0>0$ such that
$\Phi(2\alpha)\leq c\Phi(\alpha)$ for any $\alpha\geq\alpha_0$.
Thus, when $\Phi(\alpha)=|\alpha|^p$ with $p\geq1$,
we have $M^\Phi=L^\Phi=L^p$.
On the other hand, if $\Phi(\alpha)=e^{|\alpha|}-1$,
$M^\Phi$ is a proper subset of $L^\Phi$.
Moreover, if $\Phi$ takes the value $\infty$, say,
$\Phi(\alpha)=|\alpha|$ if $|\alpha|\leq1$; $=\infty$ otherwise,
then $L^\Phi=L^\infty$ and $M^\Phi=\{0\}$.
In this paper, we fix a Young function $\Phi$; and
denote by $\Psi$ its complimentary function.
Note that $L^\Psi$ is the dual space of $M^\Phi$, that is,
the set of all continuous linear functionals on $M^\Phi$.
For example, when $M^\Phi=L^p$ for $p>1$, $L^\Psi=L^{\frac{p}{p-1}}$.
Moreover, the dual space of $L^\Phi$ may include a singular part.
For more details on Orlicz space,
see Edgar and Sucheston \cite{ES92} and Rao and Ren \cite{RR}.

Let $L$ be either $M^\Phi$ or $L^\Phi$,
which is regarded as the set of all future cash-flows.
We denote by $L^*$ its dual space.
This setting would be natural,
since it covers wide classes including all $L^p$ spaces with $p\in[1,\infty]$;
and fits to utility maximization problems
(see Arai \cite{A11}, \cite{AF}, Biagini and Frittelli \cite{BF09} and
Cheridito and Li \cite{CL09}).
Moreover, let $M$ be a convex subset of $L$ including $L_-$.
The set $M$ expresses the set of all $0$-attainable claims:
future payoffs which investors can purchase without initial cost.
Although $M$ is assumed to be a convex cone in \cite{AF},
we generalize it to the convex case by excluding the cone property from $M$.

For later use, we prepare some notation.

\begin{defn}
\label{def-Q}
\begin{enumerate}
\item $\calP:=\{Q\ll\bbP|\d Q/\d\bbP\in L^\Psi\}$,
\item $L^*_1:=\{g\in L^*|g(1)=1,g(x)\geq0\mbox{ for any }x\in L_+\}$,
\item $\olL^*:=\{g\in L^*_1|\sup_{m\in M}g(m)<\infty\}$,
\item $\calQ:=\{Q\in\calP|\sup_{m\in M}\bbE_Q[m]<\infty\}$,
\item $\calQ^e:=\{Q\in\calQ|Q\sim\bbP\}$,
\item $\calQ_0:=\{Q\in\calQ|\sup_{m\in M}\bbE_Q[m]=0\}$.
\end{enumerate}
\end{defn}

\begin{rem}
When $M$ is a convex cone, $\sup_{m\in M}\bbE_Q[m]$ becomes either
$0$ or $\infty$ for $Q\in\calP$, that is, $\calQ$ and $\calQ_0$ coincide.
On the other hand, $\sup_{m\in M}\bbE_Q[m]$ may take a positive number
in our setting.
\end{rem}

\subsection{Examples for convex markets}
Here we introduce some examples such that $M$ is convex, but not a cone.

\begin{ex}[A simple illiquid market model]
\label{ex-conv0}
We consider a one-period binomial model in which one riskless asset
with zero interest rate and one risky asset are tradable.
For $t=0,1$, let $S_t$ be the price of the risky asset at time $t$.
We assume that $S_0\in\bbR$ and $S:=S_1-S_0$ belongs to $L$.
We take into account nonlinear illiquidity effects denoted
by a function $f:\bbR\to\bbR$.
More precisely, we assume that, for any $\alpha\in\bbR$,
it costs $\alpha S_0+f(\alpha)$ to get $\alpha$ units of the risky asset.
It seems that $f(\alpha)$ describes the extra cost
for purchasing $\alpha$ units of the risky asset.
Now, suppose that $f$ is a continuous convex function with $f(0)=0$,
non-increasing on $(-\infty,0]$ and non-decreasing on $[0,\infty)$.
For example, $f(\alpha)=e^{|\alpha|}-1$ or $f(\alpha)=\alpha^2$.
As a result, the set of all $0$-attainable claims is expressed as
\[
M=\{\alpha S-f(\alpha)|\alpha\in\bbR\}-L_+,
\]
which forms a convex subset of $L$ including $L_-$, but not necessarily a cone.
\end{ex}

\begin{ex}[Constraints on number of shares]
\label{ex-conv2}
Consider a continuous trading model with maturity $T\in(0,\infty)$.
Suppose that one riskless asset with zero interest rate
and $d$ risky assets are tradable; and
the price of the risky assets is described by an $\bbR^d$-valued
locally bounded RCLL special semimartingale $S$ defined on a complete
probability space $(\Omega, \calF, P ; \bbF=\{\calF_t\}_{t\in[0,T]})$,
where $\bbF$ is a filtration satisfying the so-called usual condition,
that is, $\bbF$ is right-continuous, $\calF_T=\calF$ and $\calF_0$
contains all null sets of $\calF$.
Let $L(S)$ be the set of all $\bbR^d$-valued $S$-integrable predictable
processes; and $G_t(\vt):=\int_0^t\vt_s\d S_s$ for any $t\in[0,T]$ and
any $\vt\in L(S)$.
Note that $\vt\in L(S)$ denotes the number of shares the investor holds; and
the process $G(\vt)$ represents the gain process induced
by a self-financing strategy $\vt$.
Now, we impose convex constraints on the set of all admissible strategies.
That is, we consider the case where the set of $0$-attainable claims is
given as
\[
M=\{G_T(\vt)|\vt\in L(S),\vt_t\in K\mbox{ for any }t\in[0,T]\}\cap L-L_+,
\]
where $K$ is a convex subset of $\bbR^d$ including $0$.
We introduce some concrete examples of $K$ as follows:
\begin{enumerate}
\item (Rectangular constraints)
      $K=[a_1,b_1]\times[a_2,b_2]\times\cdots\times[a_d,b_d]$
      for some fixed numbers $-\infty\leq a_i\leq0\leq b_i\leq\infty$,
      $i=1,2,\dots,d$.
\item (Constraints on total number of shares)
      $K=\{(h_1,\dots,h_d)\in\bbR^d|h_i\geq0\mbox{ for each }1\leq i\leq d,
      \sum_{i=1}^dh_i\leq c\}$ for some positive constant $c$.
\item (Short-sale constraints)
      $K=\{(h_1,\dots,h_d)\in\bbR^d|h_i\geq-c\mbox{ for each }1\leq i\leq d\}$
      for some positive constant $c$.
\end{enumerate}
For more details, see \cite{Cuoco}, \cite{CK92} and \cite{CK93}.
\end{ex}

\begin{ex}[Constraints on amount invested]
\label{ex-conv2-2}
We consider the same model as the previous example;
and assume that $S>0$ and $M$ is given as
\[
M=\{G_T(\vt)|\vt\in L(S),\vt_tS_{t-}\in K\mbox{ for any }t\in[0,T]\}\cap L-L_+,
\]
where $K$ is a convex subset of $\bbR^d$ including $0$, and
$\vt S_-=(\vt^1S^1_-,\dots,\vt^dS^d_-)$ represents
the amount invested in each asset.
The three examples for $K$ introduced in Example \ref{ex-conv2} are also
typical examples for the present setting.
\end{ex}

\begin{ex}[$a$-admissible]
\label{ex-conv3}
We consider the same mathematical framework as Example \ref{ex-conv2}.
Let $a$ be a positive real number.
$\vt\in L(S)$ is said to be $a$-admissible if
$G_t(\vt)\geq-a$ for any $t\in[0,T]$.
When $M$ is given as
\begin{equation}
\label{a-adm}
M=\{G_T(\vt)|\vt\in L(S)\mbox{ is $a$-admissible }\}\cap L-L_+
\end{equation}
for fixed $a>0$, it forms a convex set.
On the other hand, when $M$ is denoted by
$M=\{G_T(\vt)|\vt\in L(S)$ is $a$-admissible for some $a>0\}\cap L-L_+$,
it is a convex cone.
For more details, see Section 9 in Delbaen and Schachermayer \cite{DS06}.
\end{ex}

\begin{ex}[$W$-admissible]
\label{ex-conv4}
In the previous example, when $S$ is not necessarily locally bounded, 
$M$ defined in (\ref{a-adm}) may become $\{0\}$.
As a natural way to avoid it, we introduce $W$-admissibility.
Let $W$ be a random variable in $L$ with $W\geq1$.
$\vt\in L(S)$ is said to be $W$-admissible if
$G_t(\vt)\geq-W$ for any $t\in[0,T]$.
Then, $M=\{G_T(\vt)|\vt\in L(S)$ is $W$-admissible $\}\cap L-L_+$
formulates a convex market.
\end{ex}

\begin{ex}[Predictably convexity]
\label{pred-conv}
We introduce the predictably convexity,
which brings us an important class of models with convex constraints.
It has been undertaken by F\"ollmer and Kramkov \cite{FK97}; and discussed in
Chapter 9 of F\"ollmer and Schied \cite{FS11} for discrete time models.
See also \cite{A11}, Kl\"oppel and Schweizer \cite{KS05}.
Now, we define it as follows:
A family of semimartingales $\calS$ is said to be predictably convex
if, for any $S^1, S^2\in\calS$ and any $[0,1]$-valued predictable process $h$,
$\dint_0^\cdot h\d S^1+\dint_0^\cdot(1-h)\d S^2$ belongs to $\calS$.
For the three examples of portfolio constraints in Example \ref{ex-conv2},
their $M$s are predictably convex.
Here we consider the same continuous trading model as Example \ref{ex-conv2},
provided that $S$ is possibly nonlocally bounded.
Now, we fix an $\calF_T$-measurable random variable $W\in L$ with $W\geq1$
satisfying, for each $i=1,\dots,d$, there exists $\vt^i\in L(S^i)$ such that
$\bbP(\{\omega|\mbox{ there exists $t\in[0,T]$ such that }
\vt^i_t(\omega)=0\})=0$ and
$|\int_0^t\vt^i_s\d S^i_s|\leq W$ for any $t\in[0,T]$.
In addition, we denote
\begin{eqnarray*}
\Theta^W&:=&\{\vt\in L(S) | \mbox{there exists }c>0\mbox{ such that }
            G_t(\vt)\geq-cW \\
        &&  \mbox{for any }t\in[0,T]\},
\end{eqnarray*}
and $G(\Theta^W):=\{G(\vt)|\vt\in\Theta^W\}$.
Let $\calS$ be a predictably convex subset of $G(\Theta^W)$, and
$\Theta^{\calS}$ the corresponding subset of $\Theta^W$ to $\calS$.
That is, we can describe $\calS=\{G(\vt) | \vt\in\Theta^\calS\}$.
Now, we denote
\begin{equation}
\label{eq-pred}
M=\l\{G_T(\vt)|\vt\in\Theta^\calS\r\}-L_+,
\end{equation}
which is convex.
\end{ex}

\subsection{Convex risk measure}
We define convex risk measures and some related terminologies.
In addition, we introduce a representation result.

\begin{defn}
\begin{enumerate}
\item A $(-\infty,\infty]$-valued functional $\rho$ defined on $L$ is
      called a convex risk measure if $\rho$ satisfies,
      for any $x$, $y\in L$,
      \begin{description}
      \item[properness:]      $\rho(0) < \infty$,
      \item[monotonicity:]    $\rho(x)\geq\rho(y)$ if $x\leq y$,
      \item[cash-invariance:] $\rho(x+r)=\rho(x)-r$ for any $r\in\bbR$,
      \item[convexity:]       $\rho(\lambda x+(1-\lambda)y)
                              \leq\lambda\rho(x)+(1-\lambda)\rho(x)$
                              for any $\lambda\in[0,1]$.
      \end{description}
\item In addition, a convex risk measure $\rho$ is a coherent risk measure
      if it satisfies
      \begin{description}
      \item[positive homogeneity:] $\rho(\lambda x)=\lambda\rho(x)$ 
                                   for any $x\in L$ and any $\lambda\geq0$.
      \end{description}
\end{enumerate}
\end{defn}

\begin{defn}
\begin{enumerate}
\item Let $f$ be a $[-\infty,\infty]$-valued functional on $L$.
      \begin{enumerate}
      \item If $f(0)=0$, then $f$ is said to be normalized.
      \item $f$ is said to have the Fatou property if
            $\lim_{n\to\infty}f(-x_n)=f(-x)$
            for any increasing sequence $\{x_n\}\subset L$
            with $x_n\uparrow x$.
      \item $f$ is said to be relevant if
            $f(-z)>0$ for any $z \in L_+\setminus \{0\}$.
      \item We define the penalty function for $f$ as
            \begin{equation}
            \label{penalty}
            f^*(g):=\sup_{x\in L}\{g(-x)-f(x)\}
            \end{equation}
            for $g\in L^*_1$.
            In particular, we denote, for $Q\in\calP$,
            \begin{equation}
            \label{penalty2}
            f^*(Q):=\sup_{x\in L}\{\bbE_Q[-x]-f(x)\}.
            \end{equation}
      \end{enumerate}
\item We denote by $\calR$ the set of all normalized convex risk measures
      on $L$ with the Fatou property.
\end{enumerate}
\end{defn}

\begin{thm}[Proposition 1 of \cite{BF09}]
\label{BF}
Any $\rho\in\calR$ is represented as
\[
\rho(x)=\sup_{Q \in \calP}\{E_Q[-x]-\rho^*(Q)\}.
\]
\end{thm}

\subsection{A separating result}
We prepare a proposition, which will appear over and over again
in the sequel.
Now, we denote by $\olM$ (resp. $M^s$) the closure of $M$
in $\sigma(L,L^\Psi)$ (resp. in $\|\cdot\|$).

\begin{prop}
\label{prop-sep}
Let $B\subset L_+$ be a convex set including at least one positive constant.
\begin{enumerate}
\item
If $B$ is $\|\cdot\|$-compact and $M^s\cap B=\emptyset$,
then there exists a $g\in\olL^*$ such that
\begin{equation}
\label{eq-sep1}
\sup_{m\in M^s}g(m)<\inf_{x\in B}g(x).
\end{equation}
\item
If $B$ is $\sigma(L,L^\Psi)$-compact and $\olM\cap B=\emptyset$,
then there exists a $Q\in\calQ$ such that
\[
\sup_{m\in\olM}\bbE_Q[m]<\inf_{x\in B}\bbE_Q[x].
\]
\end{enumerate}
\end{prop}

\proof
It suffices to show only the first assertion.
By the conditions, the Hahn-Banach separating theorem implies
the existence of $g\in L^*$ satisfying (\ref{eq-sep1}).
Remark that $\sup_{m\in M^s}g(m)\geq0$ because $0\in M$.
Thus, we have $g(1)>0$, since $B$ includes at least one positive constant.
Without loss of generality, we may assume $g(1)=1$.
Moreover, since $L_-\subset M$, $g\in L^*_1$ holds true.
In addition, Definition \ref{def-Q} implies that
the LHS of (\ref{eq-sep1}) takes the value $\infty$ unless $g\in\olL^*$.
Thus, $g$ belongs to $\olL^*$.
\fin

\setcounter{equation}{0}
\section{Superhedging cost}
Superhedging cost for a claim is defined as the lowest price of the claim
which enables investors to construct an arbitrage opportunity by
selling the claim and selecting a suitable strategy from $M$.
More precisely, defining a functional $\rho^0$ on $L$ as
\begin{equation}
\label{eq-rho0}
\rho^0(x):=\inf\{r\in\bbR|\mbox{ there exists $m\in M$ such that }r+m+x\geq0\},
\end{equation}
the superhedging cost for claim $x$ is given by $\rho^0(-x)$;
and the no-arbitrage pricing bound for $x$ is given by
$[-\rho^0(x),\rho^0(-x)]$.
Note that GDVs will be defined by using $\rho^0$ in Section 4.
Thus, we investigate properties of $\rho^0$
which we will need for studying GDVs.

\begin{lem}
\label{lem-rho0}
$(\rho^0)^*(g)=\sup_{m\in M}g(m)$ for any $g\in L^*_1$,
where $(\rho^0)^*$ is the penalty function for $\rho^0$
defined in (\ref{penalty}).
\end{lem}

\proof
Since $\rho^0(-m)\leq0$ for any $m\in M$, (\ref{penalty}) implies that
$(\rho^0)^*(g)\geq\sup_{m\in M}\l\{g(m)-\rho^0(-m)\r\}\geq\sup_{m\in M}g(m)$
for any $g\in L^*_1$.
On the other hand, for any $x \in L$ with $\rho^0(x)<\infty$,
we take an $r>\rho^0(x)$ arbitrarily.
There is then an $m^{x,r}\in M$ satisfying $r+m^{x,r}+x\geq0$.
Since $g(m^{x,r})\leq\sup_{m\in M}g(m)$ for any $g\in L^*_1$,
we have $\sup_{m\in M}g(m)\geq g(-x)-r$, that is,
$\sup_{m\in M}g(m)\geq g(-x)-\rho^0(x)$.
In addition, this inequality also holds
for any $x\in L$ with $\rho^0(x)=\infty$.
Therefore, we have $\sup_{x\in L}\{g(-x)-\rho^0(x)\}\leq\sup_{m\in M}g(m)$
for any $g\in L^*_1$.
Consequently, $(\rho^0)^*(g)=\sup_{m\in M}g(m)$
for any $g\in L^*_1$.
\fin

\begin{prop}
\label{prop0}
$\olL^*\neq\emptyset$ if and only if $\rho^0$ is a convex risk measure on $L$.
\end{prop}

\proof
``only if" part: \ 
Firstly, the monotonicity and cash-invariance are obvious.
Next, we see $\rho^0>-\infty$.
Assuming that there exists an $x\in L$ with $\rho^0(x)=-\infty$,
(\ref{eq-rho0}) implies that for any $c>0$, we can find an $m^c\in M$
such that $-c+m^c+x\geq0$.
Thus, for any $g\in\olL^*$, we have $g(x)\geq c-(\rho^0)^*(g)$ for any $c>0$
by Lemma \ref{lem-rho0}, that is, $g(x)=\infty$.
This is a contradiction, so $\rho^0$ is $(-\infty, \infty]$-valued;
and has the properness because $\rho^0(0)\leq0$.
Lastly, we see the convexity of $\rho^0$.
Fix $x_1$, $x_2\in L$ and $\lambda\in[0,1]$ arbitrarily.
Now, we assume that both $\rho^0(x_1)$ and $\rho^0(x_2)$ are finite.
Otherwise, the convexity holds clearly.
Taking $r_i>\rho^0(x_i)$ for $i=1,2$ arbitrarily,
the convexity of $M$ implies
$\lambda r_1+(1-\lambda)r_2\geq\rho^0(\lambda x_1+(1-\lambda)x_2)$,
from which the convexity of $\rho^0$ follows.

``if" part: \ 
Suppose that $\olL^*=\emptyset$.
Assuming that there exists a $c>0$ with $c\notin M^s$,
Proposition \ref{prop-sep} implies the existence of $g\in\olL^*$ satisfying
$c>\sup_{m\in M^s}g(m)$, which is a contradiction.
As a result, any $c>0$ is included in $M^s$.
Now, for any $k\in\bbN$, we take an $m_k\in M$ satisfying $\|2^k-m_k\|\leq1$.
We define $x_n:=\sum_{k=1}^n|2^k-m_k|2^{-k}$ for any $n\in\bbN\cup\{\infty\}$.
Then, $x_n$ converges to $x_\infty$ a.s.; and $\{x_n\}$ is a Cauchy sequence
in $\|\cdot\|$.
Hence, Lemma \ref{lem-prop0} provides $x_\infty\in L$.
Noting that
$x_\infty\geq x_n\geq\sum_{k=1}^n(2^k-m_k)2^{-k}=n-\sum_{k=1}^nm_k2^{-k}$
for any $n\in\bbN$, and $\rho^0(-m)\leq0$ for any $m\in M$, we have
\[
\rho^0(x_\infty)\leq-n+\sum_{k=1}^n2^{-k}\rho^0(-m_k)
                      +\l(1-\sum_{k=1}^n2^{-k}\r)\rho^0(0)
                \leq-n
\]
for any $n\in\bbN$.
Consequently, $\rho^0(x_\infty)=-\infty$, which is a contradiction.
\fin

\begin{lem}
\label{lem-prop0}
Let $\{x_n\}_{n\geq1}$ be a Cauchy sequence on $(L, \|\cdot\|)$
which converges to $x_\infty$ a.s.
Then, $\{x_n\}$ converges to $x_\infty$ in $\|\cdot\|$,
that is, $x_\infty\in L$.
\end{lem}

\proof
Since $\{x_n\}$ is a Cauchy sequence,
there exists an $x^\prime_\infty\in L$ such that $\|x_n-x^\prime_\infty\|\to0$
by the completeness of $(L, \|\cdot\|)$.
In addition, Proposition 2.1.10 (6) of \cite{ES92} implies that
$x_n$ tends to $x^\prime_\infty$ in probability.
Hence, $x_\infty=x^\prime_\infty\in L$.
\fin

\begin{rem}
In the proof of Proposition \ref{prop0}, we see $x_\infty\in L$.
At first glance, it seems to be shown easier as follows:
$\|x_\infty\|=\|\sum_{k=1}^\infty|2^k-m_k|2^{-k}\|
\leq\sum_{k=1}^\infty2^{-k}\|2^k-m_k\|<\infty$, which implies $x_\infty\in L$.
However, this is not accurate.
Firstly, the former inequality is not trivial.
Besides, even if $\|x_\infty\|<\infty$, $x_\infty$ does not necessarily
belong to $L$, since $L$ may be $M^\Phi$ a proper subset of $L^\Phi$.
\end{rem}

\begin{rem}
When $M$ is a convex cone, $(\rho^0)^*$ takes the values $0$ and $\infty$ only.
Thus, $\rho^0$ is a coherent risk measure if and only if $\olL^*\neq\emptyset$.
For more details, see \cite{AF}.
\end{rem}

\begin{ex}
\label{ex0-0}
For the case where $M=[0,1]-L_+=\{x\in L|x\leq1\}$,
$\rho^0$ becomes a convex risk measure.
Indeed, $g(x):=\bbE[x]$ belongs to $\olL^*$.
On the other hand, setting $M=[0,\infty)-L_+=\{x\in L|x\vee0\in L^\infty\}$,
we have $\rho^0(0)=-\infty$, that is, $\rho^0$ is not a convex risk measure.
In this case, $\olL^*$ is empty evidently.
\end{ex}

\begin{rem}
We consider the concept of no arbitrage of the first kind,
which is weaker than the NFL and the no-free-lunch with vanishing risk.
We call $z\in L_+\backslash\{0\}$ an arbitrage of the first kind if,
for any $\ve>0$, we can find an $m\in M$ such that $\ve+m-z\geq0$.
For more details on arbitrage of the first kind, see Kardaras \cite{K12}.
We can see immediately that, for $z\in L_+\backslash\{0\}$,
it is an arbitrage of the first kind if and only if $\rho^0(-z)=0$.
In other words, there is no arbitrage of the first kind
if and only if $\rho^0$ is relevant.
\end{rem}

Now, we define a functional $\wh{\rho^0}$,
which is closely related to superhedging cost, as follows:
\begin{equation*}
\wh{\rho^0}(x) :=
\begin{cases}
\sup_{Q\in\calQ}\{\bbE_Q[-x]-(\rho^0)^*(Q)\} & \mbox{if }\calQ\neq\emptyset, \\
-\infty                                      & \mbox{ otherwise}.
\end{cases}
\end{equation*}
Note that $(\rho^0)^*(Q)$ is defined in (\ref{penalty2}).
We introduce a proposition, some lemmas and examples related to $\wh{\rho^0}$.

\begin{prop}
\label{prop0-2-1}
The following are equivalent:
\begin{enumerate}
   \item $\calQ \neq \emptyset$.
   \item $\wh{\rho^0}$ is the largest convex risk measure
         with the Fatou property less than $\rho^0$.
   \item There exists a $c\geq0$ such that $\bbP(\olm>c)<1$
         for any $\olm\in\olM$.
   \item There exists a $c>0$ such that $c\notin\olM$.
\end{enumerate}
\end{prop}

\proof
\noindent
1$\Leftrightarrow$2: \
This equivalence is the very definition of $\wh{\rho^0}$.

\noindent
1$\Rightarrow$3: \
Supposing $\calQ\neq\emptyset$ and taking a $Q\in\calQ$ arbitrarily,
we have $(\rho^0)^*(Q)\in[0,\infty)$ and
$\bbE_Q[\olm]-(\rho^0)^*(Q)\leq0$ for any $\olm\in\olM$,
that is, $Q(\olm-(\rho^0)^*(Q)\leq0)>0$ for any $\olm\in\olM$.
Hence, $\bbP(\olm>(\rho^0)^*(Q))<1$ for any $\olm\in\olM$.

\noindent
3$\Rightarrow$4: \
If $c\in\olM$ for any $c>0$, condition 3 is false,
since $\bbP(c+1>c)=1$ for any $c\geq0$.

\noindent
4$\Rightarrow$1: \
Taking a $c>0$ which is not included in $\olM$,
Proposition \ref{prop-sep} ensures that $\calQ$ is nonempty,
since $\{c\}$ is compact.
\fin

\begin{ex}
\label{ex0-2}
We illustrate an example in which $\rho^0\neq\wh{\rho^0}$ holds.
Let $\Omega=\{\omega_k;k\in\bbN\}$,
$\bbP(\{\omega_k\})>0$ for $k\in\bbN$, and
\begin{eqnarray*}
M&=&\Bigg\{\sum_{k=1}^\infty\vt_k 1_{\{\omega_k\}}|
    0\leq\vt_k\leq1\mbox{ for any }k\in\bbN, \\
 && \hspace{15mm}\vt_k=0\mbox{ except for finitely many $k$s}\Bigg\}-L_+.
\end{eqnarray*}
Since $\bbP\in\calQ$, $\wh{\rho^0}$ has the Fatou property
by Proposition \ref{prop0-2-1}.
On the other hand, letting $x_n:=\sum_{k=1}^n1_{\{\omega_k\}}$ for $n\in\bbN$,
we have $\rho^0(-x_n)=0$ for any $n\in\bbN$,
although $\rho^0(-1)=1$ and $x_n$ tends to $1$.
Thus, $\rho^0$ does not possess the Fatou property.
Another example in which $\rho^0\neq\wh{\rho^0}$ holds has been introduced in
\cite{AF}.
\end{ex}

\begin{lem}
\label{lem0-2-3}
The following are equivalent:
\begin{enumerate}
 \item $\calQ\neq\emptyset$ and $\inf_{Q\in\calQ}(\rho^0)^*(Q)=0$.
 \item $-\rho^0(x) \leq -\wh{\rho^0}(x) \leq \wh{\rho^0}(-x) \leq \rho^0(-x)$
       for any $x\in L$.
 \item $\wh{\rho^0}(0)=\rho^0(0)=0$.
 \item $\wh{\rho^0}(0)=0$.
\end{enumerate}
\end{lem}

\proof
1$\Rightarrow$2: \
Proposition \ref{prop0-2-1} yields that $\wh{\rho^0}\leq\rho^0$.
The convexity of $\wh{\rho^0}$ implies that
$\wh{\rho^0}(x)+\wh{\rho^0}(-x)\geq2\wh{\rho^0}(0)
=-2\inf_{Q\in\calQ}(\rho^0)^*(Q)=0$, from which the implication follows.

\noindent
2$\Rightarrow$3: \
Substituting $0$ for $x$,
we have $\rho^0(0)\geq\wh{\rho^0}(0)\geq-\wh{\rho^0}(0)$.
Thus, $\rho^0(0)\geq\wh{\rho^0}(0)\geq0$ holds.
In addition, (\ref{eq-rho0}) implies that $\rho^0(0)\leq0$.

\noindent
3$\Rightarrow$4: \
Obvious.

\noindent
4$\Rightarrow$1: \
By the definition of $\wh{\rho^0}$, $\calQ\neq\emptyset$ is ensured.
Moreover, $0=\wh{\rho^0}(0)=-\inf_{Q\in\calQ}(\rho^0)^*(Q)$ holds true.
\fin

\begin{ex}
\label{ex0-2-2}
The condition ``$\calQ\neq\emptyset$" does not ensure
``$\inf_{Q\in\calQ}(\rho^0)^*(Q)=0$".
We consider the following simple model:
Set $\Omega=\{\omega_1,\omega_2\}$,
and $S=1_{\{\omega_1\}}+\frac{1}{2}1_{\{\omega_2\}}$.
Note that we do not need to specify $\Phi$.
Let us consider the case where $M$ is given by
$\{\vt S|\vt\in[0,1]\}-L_+$.
In this case, we have $\calP=\calQ\neq\emptyset$ and
$\inf_{Q\in\calQ}(\rho^0)^*(Q)=\inf_{Q\in\calQ}\bbE_Q[S]=\frac{1}{2}$.
Hence, $\wh{\rho^0}(0)=-\frac{1}{2}$, that is,
$\wh{\rho^0}$ is not normalized.
\end{ex}

\begin{rem}
Condition 1 in Lemma \ref{lem0-2-3} is equivalent to $\calQ_0\neq\emptyset$
when $M$ is a convex cone.
Actually, it will play a similar role to ``$\calQ_0\neq\emptyset$"
in the convex cone case as in \cite{AF}.
\end{rem}

\begin{ex}
By Lemma \ref{lem0-2-3}, $\rho^0(0)=0$ whenever $\wh{\rho^0}(0)=0$,
while its reverse implication does not hold.
We reconsider Example \ref{ex0-2}.
Now, we assume $\rho^0(0)<0$.
Letting $r=\frac{-\rho^0(0)}{2}(>0)$, there exists an $m\in M$ such that
$m\geq r$, which means $m(\omega_k)\geq r$ for any $k\in\bbN$.
This is a contradiction.
As a result, $\rho^0(0)=0$.
Next, we calculate $\wh{\rho^0}(0)$.
Note that $\calQ\neq\emptyset$.
For any $Q\in\calQ$ and any $\ve>0$, we can find a finite set $A\subset\Omega$
satisfying $Q(A)>1-\ve$.
Thus, for any $\ve>0$, there exists an $m\in M$ such that $\bbE_Q[m]>1-\ve$.
So that, $(\rho^0)^*(Q)=1$ for any $Q\in\calQ$.
Consequently, we have $\wh{\rho^0}(0)=-1$.
\end{ex}

\begin{lem}
\label{lem0-2-4}
If there exists a $Q\in\calQ^e$ with $(\rho^0)^*(Q)=0$,
then $\wh{\rho^0}$ is relevant.
\end{lem}

\proof
Letting $Q$ be an element of $\calQ^e$ with $(\rho^0)^*(Q)=0$,
we have, for any $z\in L_+\backslash\{0\}$,
$\wh{\rho^0}(-z)\geq\bbE_Q[z]-(\rho^0)^*(Q)=\bbE_Q[z]>0$.
\fin

\begin{ex}
\label{ex0-3}
Even if $\inf_{Q\in\calQ}(\rho^0)^*(Q)=0$,
we may have $(\rho^0)^*(Q)>0$ for any $Q\in\calQ$.
Now, we construct such an example.
Set $\Omega=\{\omega_0,\omega_1,\dots\}$,
$\bbP(\{\omega_k\})=\frac{1}{2^{k+1}}$ for $k=0,1,2,\dots$.
Letting $M$ be given as $\{\vt S|\vt\in[0,1]\}-L_+$,
where $S(\omega_k)=\frac{1}{2^k}$ for any $k\in\bbN$,
we have $(\rho^0)^*(Q)=\bbE_Q[S]>0$ for any $Q\in\calQ$.
Defining a probability measure $Q_k$ for each $k\in\bbN$ as
$Q_k(\{\omega_l\})=1_{\{l=k\}}$ for $l\in\bbN\cup\{0\}$,
we have $Q_k\in\calQ$ for each $k\in\bbN$.
Then $\inf_{Q\in\calQ}(\rho^0)^*(Q)\leq
\inf_{k\in\bbN}(\rho^0)^*(Q_k)=\inf_{k\in\bbN}\frac{1}{2^k}=0$.
As a result, the condition in Lemma \ref{lem0-2-4} is stronger than
Condition 1 in Lemma \ref{lem0-2-3} in general.
\end{ex}

\begin{ex}
We consider the predictably convexity introduced in Example \ref{pred-conv};
and illustrate representations of $\calQ$, $\calQ_0$ and
$(\rho^0)^*$ for predictably convex models.
The following argument is based on Section 6 of \cite{A11}.
Now, we assume that $M$ defined in (\ref{eq-pred}) is included in $L$;
and define
\begin{eqnarray*}
\calP(\calS)
&:=&\{Q\in\calP|
    \mbox{ there exists increasing predictable process $A$ such that } \\
&&  G(\vt)-A\mbox{ is a $Q$-supermartingale for any }
    \vt\in\Theta^{\calS}\}.
\end{eqnarray*}
When $Q\in\calP(\calS)$, $G(\vt)$ is a special semimartingale under $Q$
for any $\vt\in\Theta^{\calS}$ (Lemma 6.2 of \cite{A11}).
Fixing $Q\in\calP(\calS)$, we denote by $M^\vt+A^\vt$
the canonical decomposition of $G(\vt)$ under $Q$.
Note that this decomposition depends on $Q$.
Now, we define $\calA:=\{A^\vt|\vt\in\Theta^{\calS}\}$.
In addition, for two stochastic processes $X$ and $Y$,
we define an order $\preceq$ as follows:
\[
X\preceq Y \Longleftrightarrow Y-X\mbox{ is an increasing process}.
\]
Remark that the ordered set $(\calA, \preceq)$ is directed upward
(Lemma 6.4 of \cite{A11}).
An increasing predictable process $A^\calS$ is called
an upper variation process of the ordered set $(\calA, \preceq)$
if $A^\calS$ satisfies the following two conditions:
\begin{enumerate}
 \item $A\preceq A^\calS$ for any $A\in\calA$,
 \item if an increasing predictable process $\wh{A}$ satisfies
       $A\preceq\wh{A}$ for any $A\in\calA$, then $A^\calS\preceq\wh{A}$ holds.
\end{enumerate}
The following assertions are from Theorem 6.9 and Theorem 5.2 in \cite{A11}.
\begin{enumerate}
\item
We have $\calQ=\calP(\calS)$ and
\[
(\rho^0)^*(Q)=\l\{
   \begin{array}{ll}
      \bbE_Q[\ol{A}^Q_T] <\infty, & \mbox{ if }Q\in\calP(\calS), \\
      \infty,                     & \mbox{ otherwise},
   \end{array}\r.
\]
where $\ol{A}^Q$ is an upper variation process for $Q\in\calP(\calS)$.
\item
$\calQ_0=\{Q\in \calP(\calS) | \mbox{ $G(\vt)$ is a $Q$-supermartingale for any }
         \vt\in\Theta^{\calS}\}$.
\end{enumerate}
\end{ex}

\setcounter{equation}{0}
\section{Good deal valuations}
In this section, we investigate thoroughly properties of GDVs.
Since any good deal bound is given as a subinterval of the no-arbitrage
pricing bound,
when we represent the upper and lower bounds of a good deal bound
as functionals $a$ and $b$ respectively,
we have $[b(x), a(x)]\subset[-\rho^0(x), \rho^0(-x)]$ for any $x\in L$.
Now, we define a functional $\rho$ as $\rho(-x):=a(x)$.
It is then natural that $\rho$ is a normalized convex risk measure
as discussed in \cite{AF}.
We call such a risk measure a GDV.
Its precise definition is given as follows:

\begin{defn}
\label{def-GDV}
A convex risk measure $\rho\in\calR$ is said to be a good deal valuation(GDV)
if
\begin{equation}
\label{eqGDV}
\rho(-x)\in[-\rho^0(x), \rho^0(-x)]\mbox{ for any }x\in L.
\end{equation}
\end{defn}

\noindent
Note that we consider only convex risk measures having the Fatou property
as GDVs in this paper.
Although the definition (\ref{eqGDV}) is given from the seller's view point,
we can rewrite (\ref{eqGDV}) as
\begin{equation}
\label{eqGDV-2}
-\rho(x)\in[-\rho^0(x), \rho^0(-x)]\mbox{  for any }x\in L,
\end{equation}
which means that any GDV describes the lower bound of a good deal bound.
Indeed, denoting $-\rho^b(x):=b(x)$, $\rho^b$ satisfies (\ref{eqGDV-2}).
Furthermore, note that any GDV $\rho$ satisfies $-\rho(x)\leq\rho(-x)$
for any $x\in L$ since $\rho(x)+ \rho(-x)\geq 2 \rho(0)=0$ by the convexity.
Then, for any GDV $\rho$, the interval $[-\rho(x), \rho(-x)]$
provides a good deal bound.
Note that the upper and lower bounds of a good deal bound are mostly
described by different GDVs.

Now, we show equivalent conditions for the existence of a GDV.

\begin{thm}
\label{thm1}
The following are equivalent:
\begin{enumerate}
\item $\calQ\neq\emptyset$ and $\inf_{Q\in\calQ}(\rho^0)^*(Q)=0$.
\item $\wh{\rho^0}$ is a GDV.
\item There exists a GDV.
\item $\bbP(\olm>\ve)<1$ for any $\ve>0$ and any $\olm\in\olM$.
\item $c\notin\olM$ for any $c>0$.
\end{enumerate}
\end{thm}

\proof
1$\Rightarrow$2:
By Proposition \ref{prop0-2-1} and Lemma \ref{lem0-2-3}.

\noindent
2$\Rightarrow$3:
Obvious.

\noindent
3$\Rightarrow$1:
Let $\rho$ be a GDV.
Since $\rho(-m)\leq\rho^0(-m)\leq0$ for any $m\in M$, we have
\begin{eqnarray}
\label{eq5to1}
\rho^*(Q)&=&    \sup_{x\in L}\{\bbE_Q[-x]-\rho(x)\}
                \geq\sup_{m\in M}\{\bbE_Q[m]-\rho(-m)\} \nonumber \\
         &\geq& \sup_{m\in M}\bbE_Q[m]
                =(\rho^0)^*(Q).
\end{eqnarray}
Thus, $\rho^*(Q)=\infty$ for any $Q\in\calP\backslash\calQ$.
Supposing $\calQ=\emptyset$,
$\rho$ equals to $-\infty$ identically by Theorem \ref{BF}.
This is a contradiction.
In addition, we have
$0\leq\inf_{Q\in\calQ}(\rho^0)^*(Q)\leq\inf_{Q\in\calQ}\rho^*(Q)=0$
since $\rho(0)=0$.

\noindent
1$\Rightarrow$4:
Supposing that there exist an $\ve>0$ and an $\olm\in\olM$
such that $\bbP(\olm>\ve)=1$,
we have $\bbE_Q[\olm]>\ve$ for any $Q\in\calP$.
That is, $\sup_{m\in M}\bbE_Q[m]=\sup_{\olm\in\olM}\bbE_Q[\olm]>\ve$
for any $Q\in\calP$.
From the view of Lemma \ref{lem-rho0}, either $\calQ=\emptyset$ or
$\inf_{Q\in\calQ}(\rho^0)^*(Q)>0$ holds true.

\noindent
4$\Rightarrow$5:
We can see this by contraposition.

\noindent
5$\Rightarrow$1:
We fix $c>0$ arbitrarily.
Since $c\notin\olM$, 
Proposition \ref{prop-sep} implies that there exists a $Q_c\in \calQ$
such that $(\rho^0)^*(Q_c)=\sup_{\olm\in\olM}\bbE_{Q_c}[\olm]<c$.
By the arbitrariness of $c>0$, we have $\inf_{Q\in\calQ}(\rho^0)^*(Q)=0$.
\fin

\begin{rem}
\begin{enumerate}
\item As seen in Example \ref{ex0-3},
      even if there exists a GDV, we may find an $m\in M$
      such that $\bbP(m>0)=1$, that is, an arbitrage opportunity
      in a strong sense.
      In other words, all conditions in Theorem \ref{thm1} are not sufficient
      for the no-arbitrage condition.
\item The condition $\calQ\neq\emptyset$ is not sufficient for $\wh{\rho^0}$
      to be a GDV, since it is not necessarily normalized.
      See Example \ref{ex0-2-2}.
\item The first condition in Theorem \ref{thm1} is stronger than
      $\olL^*\neq\emptyset$.
      That is, $\rho^0$ is not necessarily a GDV even if $\olL^*\neq\emptyset$.
      See Example \ref{ex0-2}.
\end{enumerate}
\end{rem}

\begin{rem}
Theorem 3.2 of \cite{AF} provided equivalent conditions
for the existence of a GDV when $M$ is a convex cone.
Now, we shall compare Theorem \ref{thm1} with it.
\begin{enumerate}
\item The third condition of Theorem 3.2 in \cite{AF}:
      ``$\bbP(\olm>0)<1$ for any $\olm\in\olM$" is sufficient,
      but not necessary for the existence of a GDV
      in our setting as seen in Example \ref{ex0-3}.
\item The fourth condition in \cite{AF}: ``$1\notin\olM$"
      is equivalent to condition 5 in Theorem \ref{thm1}
      when $M$ is a convex cone,
      whereas condition 5 is stronger than ``$1\notin\olM$"
      unless $M$ is a cone.
\end{enumerate}
\end{rem}

Next, we enumerate equivalent conditions for a given $\rho\in\calR$
to be a GDV.

\begin{prop}
\label{prop2-1}
Let $\rho\in\calR$.
The following are equivalent:
\begin{enumerate}
 \item $\rho$ is a GDV.
 \item $\rho(-m)\leq0$ for any $m\in M$.
 \item $\rho^*(Q)\geq(\rho^0)^*(Q)$ for any $Q\in\calP$, that is,
       $\rho$ is represented as
       \[
       \rho(x)=\sup_{Q\in\calQ}\{\bbE_Q[-x]-\rho^*(Q)\}.
       \]
 \item $\rho(-x)\in[-\wh{\rho^0}(x),\wh{\rho^0}(-x)]$ for any $x\in L$.
 \item $\{\rho^0 \leq 0\} \subset \{\rho \leq 0\}$.
\end{enumerate}
\end{prop}

\proof
1$\Rightarrow $2:
For any $m\in M$, we have $\rho(-m)\leq\rho^0(-m)\leq0$ by (\ref{eq-rho0}).

\noindent
2$\Rightarrow $3:
This is from (\ref{eq5to1}).

\noindent
3$\Rightarrow $4:
For any $x\in L$, we have
\[
\rho(x)=    \sup_{Q\in\calQ}\l\{\bbE_Q[-x]-\rho^*(Q)\r\}
        \leq \sup_{Q\in\calQ}\l\{\bbE_Q[-x]-(\rho^0)^*(Q)\r\}
        =    \wh{\rho^0}(x).
\]
Moreover, the convexity of $\rho$ yields that
$-\rho(-x)\leq\rho(x)\leq\wh{\rho^0}(x)$.

\noindent
4$\Rightarrow $1:
Note that $\calQ\neq\emptyset$ holds under condition 4.
Thus, $\wh{\rho^0}\leq\rho^0$ by Proposition \ref{prop0-2-1}.
In addition, $\wh{\rho^0}(0)\geq0$ since $\rho(0)=0$.
So that, $\rho^0(0)=\wh{\rho^0}(0)=0$ because $\rho^0(0)\leq0$.
As a result, Lemma \ref{lem0-2-3} ensures that $\rho$ is a GDV.

\noindent
3$\Rightarrow $5:
Recall that $\calQ\neq\emptyset$ is ensured under condition 3.
We have
\begin{eqnarray*}
\rho^0(x)\leq0
&\Rightarrow& \mbox{ for any }\ve>0,\mbox{ there exists }m\in M
              \mbox{ such that }\ve+m+x\geq0 \\
&\Rightarrow& \mbox{ for any }\ve>0, \ve+\bbE_Q[x]+(\rho^0)^*(Q)\geq0
              \mbox{ for any }Q\in\calQ \\
&\Rightarrow& \mbox{ for any }\ve>0, \ve+\bbE_Q[x]+\rho^*(Q)\geq0
              \mbox{ for any }Q\in\calQ \\
&\Rightarrow& \rho(x)\leq\ve\mbox{ for any }\ve>0 \\
&\Rightarrow& \rho(x)\leq0.
\end{eqnarray*}

\noindent
5$\Rightarrow $2:
Remark that we have $\rho^0(-m)\leq0$ for any $m\in M$.
Thus, $-m\in\{\rho\leq0\}$ for any $m\in M$.
\fin

\subsection{Relationship with risk indifference price}
When $M$ is a convex cone,
$\rho\in\calR$ is a GDV if and only if it is a risk indifference price,
as shown in Theorem 3.4 of \cite{AF}.
However, we cannot generalize this result to our setting.
In this subsection, we investigate relationship between GDVs and
risk indifference prices.
We start with the definition of risk indifference prices.

\begin{defn}
For a given $[-\infty,\infty]$-valued functional $f$ on $L$, 
we define a functional $I(f)$  on $L$ as
\[
I(f)(x):=\inf\l\{r\in\bbR|\inf_{m\in M}f(r+m+x)
         \leq\inf_{m\in M}f(m)\r\}.
\]
In particular, when $\rho$ is a convex risk measure,
$I(\rho)$ is said the risk indifference price induced by $\rho$; and
is represented as
\[
I(\rho)(x)
=\inf\l\{r\in\bbR|\inf_{m\in M}\rho(m+x)-r\leq\inf_{m\in M}\rho(m)\r\}.
\]
\end{defn}

\noindent
$I(\rho)(-x)$ describes the risk indifference seller's price for $x$
induced by $\rho$ as introduced in Xu \cite{X06}.
Selling $x$ for a price greater than $I(\rho)(-x)$,
the investor can find a suitable strategy from $M$ so that
the risk measured by $\rho$ does not increase.
For more details on $I(\rho)$, see \cite{AF}, \cite{KS05} and \cite{X06}.
Now, we prepare a lemma as follows:

\begin{lem}
\label{implicit}
Let $\rho$ be a convex risk measure on $L$.
If $I(\rho)$ is $(-\infty, \infty]$-valued,
then we have $\inf_{m\in M}\rho(m)\in\bbR$ and that
$I(\rho)$ is a convex risk measure with
\begin{equation*}
I(\rho)^*(g)
=
\begin{cases}
(\rho^0)^*(g)+\rho^*(g)+\inf_{m\in M}\rho(m), & \mbox{ if }g \in \olL^*, \\
\infty,                                       & \mbox{ otherwise}.
\end{cases}
\end{equation*}
If $I(\rho)\in\calR$ in addition, then $\calQ \neq \emptyset$ and
\[
I(\rho)(x)=\sup_{Q\in\calQ}\l\{\bbE_Q[-x]
           -(\rho^0)^*(Q)-\rho^*(Q)-\inf_{m\in M}\rho(m)\r\}.
\]
\end{lem}

\proof
We can see the lemma by the same way as the proof of Lemma 2.10 in \cite{AF}
together with the above Lemma \ref{lem-rho0}.
\fin

We illustrate an example of a GDV which is not a risk indifference price; and
two examples of GDVs which are risk indifference prices.

\begin{ex}
\label{ex-ill1}
We consider a simple illiquid market model as introduced in
Example \ref{ex-conv0}.
Let $\Omega=\{\omega_1, \omega_2\}$, $\bbP(\{\omega_i\})>0$ for $i=1,2$, and
\[
S(\omega)=\l\{
          \begin{array}{ll}
          1  & \mbox{if }\omega=\omega_1, \\
          -1 & \mbox{if }\omega=\omega_2.
          \end{array}\r.
\]
The set of $0$-attainable claims is given as
\[
M=\{\alpha S-\alpha^2|\alpha\in\bbR\}-L_+
 =\{\alpha S-\alpha^2|\alpha\in[-1/2,1/2]\}-L_+.
\]
For a probability measure $Q$, we denote $q:=Q(\{\omega_1\})$,
and identify $Q$ with $q$.
Thus, we regard $[0,1]$ as the set of all probability measures.
We have
\[
(\rho^0)^*(q)
=\sup_{\alpha\in[-\frac{1}{2},\frac{1}{2}]}\l\{\bbE_q[\alpha S]-\alpha^2\r\} 
=\sup_{\alpha\in[-\frac{1}{2},\frac{1}{2}]}\alpha(2q-1-\alpha)
=\frac{(2q-1)^2}{4}.
\]
Next, we define
\[
\rho(x)=\sup_{q\in[0,1]}\l\{\bbE_q[-x]-\frac{|2q-1|}{4}\r\}.
\]
Since $\rho(x)\leq0$ whenever $\wh{\rho^0}(x)\leq0$,
Proposition \ref{prop2-1} implies that $\rho$ is a GDV.
Moreover, for $x=-\frac{1}{2}1_{\{\omega_1\}}$,
$\rho(x)=\frac{1}{4}$ and $\wh{\rho^0}(x)=\frac{5}{16}$,
namely, $\rho\neq\wh{\rho^0}$.

We show that $\rho$ is not a risk indifference price.
Suppose that $\rho$ is represented as $\rho=I(\eta)$
for some convex risk measure $\eta$.
Since $(\rho^0)^*(q)\leq\rho^*(q)\leq\frac{|2q-1|}{4}$
by Proposition \ref{prop2-1} and (\ref{penalty2}),
we have $\rho^*(q)=(\rho^0)^*(q)$ for $q\in\{0,\frac{1}{2},1\}$.
In addition, Lemma \ref{implicit} implies that
$\rho^*(q)=I(\eta)^*(q)=(\rho^0)^*(q)+\eta^*(q)+\inf_{m\in M}\eta(m)$.
Thus, $\eta^*(q)+\inf_{m\in M}\eta(m)=0$ for $q\in\{0,\frac{1}{2},1\}$.
Hence, the convexity of $\eta^*$ implies that
$\eta^*(q)+\inf_{m\in M}\eta(m)=0$ for any $q\in[0,1]$, that is,
$\rho^*=(\rho^0)^*$, which is a contradiction.
\fin
\end{ex}

\begin{ex}[Exponential utility indifference price]
For $\gamma>0$, we set $\Phi(\alpha)=e^{\gamma|\alpha|}-1$ and $L=M^\Phi$.
For an agent having an initial capital $c\in\bbR$ and
an exponential utility function with risk-aversion $\gamma$,
the utility indifference seller's price $p(-x)$ for $x\in L$
is defined implicitly as
\[
\sup_{m\in M}\bbE[-\exp\{-\gamma(c+m)\}]
=\sup_{m\in M}\bbE[-\exp\{-\gamma(c+p(-x)+m-x)\}].
\]
For more details, see Biagini et al. \cite{BFG}.
Denoting $\rho_\gamma(x):=\frac{1}{\gamma}\log\bbE[\exp\{-\gamma x\}]$,
we have $p(x)=I(\rho_\gamma)(x)$.
Note that $\rho_\gamma$ is called an entropic risk measure.
Assuming $\bbE[m]\leq0$ for any $m\in M$ additionally,
we have $\inf_{m\in M}\rho_\gamma(m)=0$, that is,
$p(-m)\leq0$ for any $m\in M$.
Hence, $p$ is a GDV.
\fin
\end{ex}

\begin{ex}[Shortfall risk measure]
We consider an agent selling a claim $x$ with price $r\in\bbR$;
and selecting $m\in M$ as her strategy.
Her shortfall risk is then defined as a weighted expectation of
the shortfall of her final cash-flow $r+m-x$ with a loss function $l$.
Note that $l$ represents her attitude towards risk.
Now, we assume that $l$ is given as $l(\alpha)=\Phi(0\wedge\alpha)$; and
$L=M^\Phi$.
For simplicity, we assume the continuity of $l$.
To suppress the shortfall risk less than a certain level $\delta>0$
which she can endure, the least price she can accept is given as
\[
\rho_l(-x):=\inf\{r\in\bbR|\mbox{ there exists }m\in M
            \mbox{ such that }\bbE[l(r+m-x)]\leq\delta\}.
\]
As seen in Arai \cite{A11}, $\rho_l$ is a convex risk measure
with the Fatou property under mild conditions.
We define $\wh{\rho_l}$ as $\wh{\rho_l}(x):=\rho_l(x)-\rho_l(0)$.
Denoting $\rho^1_l(x):=\inf\{r\in\bbR|\bbE[l(r+x)]\leq\delta\}$,
we have $\wh{\rho_l}=I(\rho^1_l)$.
As seen in the previous example,
supposing $\bbE[m]\leq0$ for any $m\in M$,
we have $\inf_{m\in M}\rho^1_l(m)=\rho^1_l(0)$,
from which $\wh{\rho_l}$ is a GDV.
\fin
\end{ex}

As seen in Example \ref{ex-ill1}, a GDV is not necessarily a risk indifference
price.
Accordingly, the following theorem gives sufficient conditions
for a GDV to be a risk indifference price; and
for a risk indifference price to be a GDV.

\begin{thm}
\label{thm2}
Let $\rho\in\calR$.
We consider the following conditions:
\begin{enumerate}
 \item There exists an $\eta\in\calR$
       with $\inf_{m\in M}\eta(m)=0$ such that $\rho=I(\eta)$.
 \item[1$^\prime$.] There exists a convex risk measure $\eta$
       with $\eta(0)=\inf_{m\in M}\eta(m)$ such that $\rho=I(\eta)$
 \item There exists a convex set $A\subset L$ including $0$
       with $A+L_+\subset A$ such that for any $x\in L$
       \begin{equation}
       \label{eq-thm2-2}
       \rho(x)=\inf\{r\in\bbR|\mbox{there exists }m\in M\mbox{ such that }
               r+m+x\in A\}.
       \end{equation}
 \item $\rho$ is a GDV.
\end{enumerate}
Then 1$\Rightarrow $1$^\prime\Leftrightarrow$2$\Rightarrow$3 holds.
Moreover, when $\rho^*-(\rho^0)^*$ is convex and $M$ is given as $M=M_0-L_+$
for some $\sigma(L,L^\Psi)$-compact convex set $M_0$ including $0$,
all the above conditions are equivalent.
\end{thm}

\proof
1$\Rightarrow $1$^\prime$ is obvious.

\noindent
1$^\prime\Rightarrow $2:
Denoting $\eta^\prime:=\eta-\inf_{m\in M}\eta(m)$,
we have $\eta^\prime(0)=\inf_{m\in M}\eta^\prime(m)=0$.
Let $A:=\{x\in L|\eta^\prime(x)\leq0\}$ and $A_\rho:=\{x\in L|\rho(x)\leq0\}$.
Note that $A$ is a convex set including $0$ with $A+L_+\subset A$.
We have
\[
\{x\in L|\mbox{ there exists $m^\prime\in M$ such that $m^\prime+x\in A$}\}
\subset A_\rho,
\]
since $\rho(x)
=I(\eta)(x)=\inf_{m\in M}\eta^\prime(m+x)\leq\eta^\prime(m^\prime+x)\leq0$
if $x$ belongs to the LHS.
Thus, we have
\begin{eqnarray*}
\rho(x)
&=&    \inf\{r\in\bbR|x+r\in A_\rho\} \\
&\leq& \inf\{r\in\bbR|\mbox{there exists }m\in M\mbox{ such that }m+x+r\in A\}.
\end{eqnarray*}
As for the reverse inequality, we have, for any $\ve>0$,
\begin{eqnarray*}
\rho(x)
&=&    \inf\{r\in\bbR|\inf_{m\in M}\eta^\prime(m+x)\leq r\} \\
&\geq& \inf\{r\in\bbR|\mbox{there exists }m\in M\mbox{ such that }
       \eta^\prime(m+x)\leq r+\ve\} \\
&=&    \inf\{r\in\bbR|\mbox{there exists }m\in M\mbox{ such that }
       m+x+r\in A\}-\ve.
\end{eqnarray*}
By the arbitrariness of $\ve$, we obtain (\ref{eq-thm2-2}).

\noindent
2$\Rightarrow $1$^\prime$:
Denote $\eta(x):=\inf\{r\in\bbR|r+x\in A\}$.
Noting that $\eta>-\infty$ by $\eta\geq\rho$; and $\eta(0)=0$ by $0\in A$,
we obtain that $\eta$ is a normalized convex risk measure
by the conditions on $A$.
Hence, it suffices to see
\begin{equation}
\label{eq-000}
\inf_{m\in M}\eta(m+x)=\rho(x),
\end{equation}
since $\inf_{m\in M}\eta(m)=0$ holds if (\ref{eq-000}) holds.
Remark that $\inf_{m\in M}\eta(m+x)=\infty \Leftrightarrow$
$r+m+x\notin A$ for any $r\in\bbR$ and any $m\in M$
$\Leftrightarrow \rho(x)=\infty$.
Then, we suppose that both $\inf_{m\in M}\eta(m+x)$ and $\rho(x)$
are less than $\infty$.
For any $r>\inf_{m\in M}\eta(m+x)$,
there exists an $m\in M$ such that $r+m+x\in A$.
Thus, $\rho(x)\leq r$.
On the other hand, for any $r>\rho(x)$, there exists an $m\in M$ such that
$r+m+x\in A$, that is, $\eta(m+x)\leq r$, which implies that
$\inf_{m\in M}\eta(m+x)\leq r$.
As a result, we have (\ref{eq-000}).

\noindent
2$\Rightarrow $3:
As seen in the above, $\rho=I(\eta)$ holds under condition 2.
Then, Lemma \ref{implicit} provides that $\rho^*=I(\eta)^*=\eta^*+(\rho^0)^*$.
Since $\eta^*\geq0$ by $\eta(0)=0$,
we have $\rho^*\geq(\rho^0)^*$.
Proposition \ref{prop2-1} implies that $\rho$ is a GDV.

As for the second assertion, it suffices to see
the implication 3$\Rightarrow $1.
Define $\wt{\rho}(x):=\sup_{Q\in\calQ}\{\bbE_Q[-x]-\rho^*(Q)+(\rho^0)^*(Q)\}$.
Since $\wt{\rho}\geq\rho>-\infty$ and $\wt{\rho}(0)\leq0$
by Proposition \ref{prop2-1},
$\wt{\rho}$ is a convex risk measure with the Fatou property.
Remark that $\wt{\rho}(m)\geq\sup_{Q\in\calQ}\{-\rho^*(Q)\}=\rho(0)=0$
for any $m\in M$, that is, $\wt{\rho}(0)=0$ and $\inf_{m\in M}\wt{\rho}(m)=0$.
Thus, we have
\begin{eqnarray*}
I(\wt{\rho})(x)
&=&\inf_{m\in M}\wt{\rho}(m+x)-\inf_{m\in M}\wt{\rho}(m)
   =\inf_{m\in M_0}\wt{\rho}(m+x) \\
&=&\inf_{m\in M_0}\sup_{Q\in\calQ}\{\bbE_Q[-m-x]-\rho^*(Q)+(\rho^0)^*(Q)\} \\
&=&\sup_{Q\in\calQ}\inf_{m\in M_0}\{\bbE_Q[-m-x]-\rho^*(Q)+(\rho^0)^*(Q)\} \\
&=&\sup_{Q\in\calQ}\{\bbE_Q[-x]-\rho^*(Q)\}=\rho(x),
\end{eqnarray*}
since the minimax theorem (Theorem 3.1 of Simons \cite{Simons}) is applicable
by the compactness of $M_0$ and the convexity of $\rho^*-(\rho^0)^*$.
\fin

\begin{rem}
\begin{enumerate}
\item 
Theorem 3.4 in \cite{AF} asserts that, when $M$ is a convex cone,
the following are equivalent for $\rho\in\calR$:
(a) $\rho$ is a GDV; (b) there exists $\eta\in\calR$ such that $\rho=I(\eta)$;
and (c) condition 2 in Theorem \ref{thm2}.
Now, recall that $\inf_{m\in M}\eta(m)=0$ automatically holds
in the convex cone markets.
That's because condition 1 in Theorem \ref{thm2} is stronger than
the above condition (b).

\item 
When $M$ is a convex cone, $\wt{\rho}$ coincides with $\rho$;
and $\inf_{m\in M}\rho(m)=0$ holds.
Thus, we do not need the minimax theorem to see
the implication 3$\Rightarrow $1 in the convex cone case.

\item
Madan and Cherny \cite{MC} developed a theory for bid and ask prices.
They gave a framework of bid and ask prices which are expressed
in a similar way with (\ref{eq-thm2-2}),
employing the concept of acceptability indexes and acceptability levels.
\end{enumerate}
\end{rem}

\subsection{Extension to conical market}
Here we consider a conical market generated by
the convex constrained market $M$.
We define a convex cone set generated by $M$ as
\[
M^\prime:=\{cm|c\geq0, m\in M\};
\]
and regard it as the set of all $0$-attainable claims in the extended market.
Now, for a given $\rho\in\calR$, we denote
\[
\rho^\prime(x):=\sup_{Q\in\calQ_0}\l\{\bbE_Q[-x]-\rho^*(Q)\r\}.
\]
Note that $\rho^\prime$ is a convex risk measure on $L$
with the Fatou property whenever $\calQ_0\neq\emptyset$, and vice versa.
In addition, $\rho^\prime\in\calR$ if and only if
$\inf_{Q\in\calQ_0}\rho^*(Q)=0$.
We show the following proposition:

\begin{prop}
For any GDV $\rho$ (for the market $M$), if $\rho^\prime\in\calR$,
then $\rho^\prime$ is the largest GDV
for the extended conical market $M^\prime$ smaller than $\rho$.
\end{prop}

\proof
Since $\bbE_Q[m^\prime]\leq0$ for any $m^\prime\in M^\prime$ and $Q\in\calQ_0$,
we have
\[
\rho^\prime(-m^\prime)
=    \sup_{Q\in\calQ_0}\l\{\bbE_Q[m^\prime]-\rho^*(Q)\r\}
\leq \sup_{Q\in\calQ_0}\l\{-\rho^*(Q)\r\}
=    0
\]
for any $m^\prime\in M^\prime$, which means that $\rho^\prime$ is a GDV
for $M^\prime$ by Proposition \ref{prop2-1}.

Now, $\rho^\prime$ is smaller than $\rho$, that is,
$\rho^\prime(x)\leq\rho(x)$ for any $x\in L$.
Taking $\rho_1$ a GDV for $M^\prime$ smaller than $\rho$ arbitrarily,
we show $\rho^\prime\geq\rho_1$.
Denoting by $\rho_1^*$ the penalty function of $\rho_1$,
we have 
$
\rho_1^*(Q)
=    \sup_{x\in L}\l\{\bbE_Q[x]-\rho_1(-x)\r\}
\geq \sup_{x\in L}\l\{\bbE_Q[x]-\rho(-x)\r\}
=    \rho^*(Q)
$
for any $Q\in\calQ$.
Note that, for any $Q\notin\calQ_0$, there exists an $m^\prime_1\in M^\prime$
such that $\bbE_Q[m^\prime_1]>0$, that is,
$\sup_{m^\prime\in M^\prime}\bbE_Q[m^\prime]=\infty$
by the cone property of $M^\prime$.
Hence, for any $Q\in\calQ\backslash\calQ_0$, we have
\[
\rho_1^*(Q)
\geq\sup_{m^\prime\in M^\prime}\l\{\bbE_Q[m^\prime]-\rho_1(-m^\prime)\r\}
\geq\sup_{m^\prime\in M^\prime}\bbE_Q[m^\prime]=\infty.
\]
Consequently, we obtain
\[
\rho^\prime(x)=\sup_{Q\in\calQ_0}\l\{\bbE_Q[-x]-\rho^*(Q)\r\}
\geq\sup_{Q\in\calQ_0}\l\{\bbE_Q[-x]-\rho_1^*(Q)\r\}=\rho_1(x)
\]
for any $x\in L$.
\fin

\subsection{Coherent good deal valuations}
When $M$ is a convex cone, $\wh{\rho^0}$ is coherent, that is,
there is a coherent GDV whenever a GDV exists.
On the other hand, in our setting,
since $\wh{\rho^0}$ is not necessarily coherent,
there might be no coherent GDV even if a GDV exists.
Now, we illustrate an equivalent condition for the existence of a coherent GDV.

\begin{prop}
\label{prop-coh}
$\calQ_0\neq\emptyset$ if and only if there exists a coherent GDV.
\end{prop}

\proof
Suppose $\calQ_0\neq\emptyset$.
Taking a $Q\in\calQ_0$,
we define $\rho_Q(x):=\bbE_Q[-x]$ for any $x\in L$.
Note that $\rho_Q$ is in $\calR$ and coherent.
We have then
$\rho_Q(-m)=\bbE_Q[m]\leq\sup_{m\in M}\bbE_Q[m]=0$ for any $m\in M$,
from which $\rho_Q$ is a GDV.

To see the reverse implication, let $\rho$ be a coherent GDV.
Since $\rho$ is coherent, $\rho^*$ takes the values $0$ and $\infty$ only.
Defining $\wt{\calQ}:=\{Q\in\calQ|\rho^*(Q)=0\}$,
we have that $\wt{\calQ}$ is nonempty and
$\rho(x)=\sup_{Q\in\wt{\calQ} }\bbE_Q[-x]$.
Proposition \ref{prop2-1} implies that,
for any $m\in M$ and any $\wt{Q}\in\wt{\calQ}$,
$0\geq\rho(-m)=\sup_{Q\in\wt{\calQ} }\bbE_Q[m]\geq\bbE_{\wt{Q} }[m]$.
Thus, $\sup_{m\in M}\bbE_{\wt{Q} }[m]=0$ for any $\wt{Q}\in\wt{\calQ}$,
that is, $\wt{\calQ}\subset\calQ_0$.
\fin

\setcounter{equation}{0}
\section{Fundamental Theorem of Asset Pricing}
In this section, we prove a Kreps-Yan type FTAP with convex constraints.
Basically, the Kreps-Yan theorem (\cite{K81} or Section 5 in \cite{DS06})
asserts, very roughly speaking, the equivalence between the existence
of an equivalent martingale measure and the NFL: $\olM\cap L_+=\{0\}$.
\cite{AF} proved, for the case where $M$ is a convex cone,
the equivalence among the NFL, $\calQ_0\cap\calQ^e\neq\emptyset$ and
the existence of a relevant GDV.
Noting that $\calQ$ and $\calQ_0$ coincide when $M$ is a convex cone,
and taking Theorem \ref{thm1} and Lemma \ref{lem0-2-4} into account,
we naturally expect the equivalence between the NFL and
either condition 1 or 1$^\prime$ of the following theorem,
whereas neither of them actually holds.
On the other hand, the equivalence between the NFL and the existence of
a relevant GDV still holds.
The following is an FTAP for markets with convex constraints:

\begin{thm}
\label{thm3}
As for the following conditions,
we have 1$^\prime\Rightarrow$4$\Leftrightarrow$3$\Leftrightarrow
$2$\Rightarrow$1.
\begin{enumerate}
\item $\calQ^e\neq\emptyset$ and $\inf_{Q\in\calQ^e}(\rho^0)^*(Q)=0$.
\item[1$^\prime$.] There exists a $Q\in\calQ^e$ with $(\rho^0)^*(Q)=0$.
\item $\olM\cap L_+=\{0\}$.
\item There exists a relevant GDV.
\item $\wh{\rho^0}$ is a relevant GDV.
\end{enumerate}
\end{thm}

\proof
2$\Rightarrow$1:
For each $\delta\in(0,1]$, we define a set $B_\delta$ as
\begin{equation}
\label{B-delta}
B_\delta:=\{x\in L|0\leq x\leq1,\bbE[x]\geq\delta\}.
\end{equation}
Note that $B_\delta$ is compact in $\sigma(L,L^\Psi)$
and $\olM\cap B_\delta=\emptyset$.
Thus, Proposition \ref{prop-sep} ensures
the existence of $Q_\delta\in \calQ$ satisfying
\begin{equation}
\label{eq-FTAP1}
\sup_{\olm\in\olM}\bbE_{Q_\delta}[\olm]
     <\inf_{x\in B_\delta}\bbE_{Q_\delta}[x].
\end{equation}
Now, we denote $Q^{(k)}:=Q_{2^{-k}}\in\calQ$
($Q_{2^{-k}}$ is defined in (\ref{eq-FTAP1}) for $\delta=2^{-k}$)
for any $k\in\bbN$; $\alpha_k:=\l\|\frac{\d Q^{(k)}}{\d\bbP}\r\|_\Psi\vee1$
($\|y\|_\Psi:=\inf\{c>0|\bbE[\Psi(y/c)]\leq1\}$); and
$C_n:=\sum_{k=n}^\infty\frac{2^{-k}}{\alpha_k}<\infty$ for any $n\in\bbN$.
Moreover, we define $\beta^n_k:=\frac{2^{-k}}{C_n\alpha_k}$ for any $k\geq n$;
and $\wt{Q}^{(n)}:=\sum_{k=n}^\infty\beta^n_kQ^{(k)}$ for any $n\in\bbN$.
Note that $\wt{Q}^{(n)}$ is a probability measure equivalent to $\bbP$,
since $\sum_{k=n}^\infty\beta^n_k=1$ and
$Q^{(k)}(A)>0$ for any $A\in\calF$ with $\bbP(A)>2^{-k}$ by (\ref{eq-FTAP1}).
Now, we denote $\gamma_i:=\sum_{k=n}^{n+i}\beta^n_k\frac{\d Q^{(k)}}{\d\bbP}$
for $i=1,2,\dots$.
Then, $\{\gamma_i\}$ is a Cauchy sequence in $\|\cdot\|_\Psi$;
and Lemma \ref{lem-prop0} yields $\frac{\d\wt{Q}^{(n)} }{\d\bbP}\in L^\Psi$.
Moreover, noting that $2^{-n}\in B_{2^{-k}}$ for any $k\geq n$,
we have, for any $n\in\bbN$,
\begin{eqnarray*}
\sup_{\olm\in\olM}E_{\wt{Q}^{(n)} }[\olm]
&=&    \sup_{\olm\in\olM}\sum_{k=n}^\infty\beta^n_kE_{Q^{(k)}}[\olm]
       \leq \sum_{k=n}^\infty\beta^n_k
            \sup_{\olm\in\olM}E_{Q^{(k)}}[\olm] \\
&<&    \sum_{k=n}^\infty\beta^n_k\inf_{x\in B_{2^{-k}}}E_{Q^{(k)}}[x]
       \leq\sum_{k=n}^\infty\beta^n_k2^{-n}=2^{-n},
\end{eqnarray*}
which implies $\wt{Q}^{(n)}\in\calQ^e$ with $(\rho^0)^*(\wt{Q}^{(n)})<2^{-n}$.
As a result, we obtain $\inf_{Q\in \calQ^e}(\rho^0)^*(Q)=0$.

\noindent
4$\Rightarrow$3:
Obvious.

\noindent
3$\Rightarrow$2:
Let $\rho$ be a relevant GDV.
Since $\rho(-z) > 0$ for all $z \in L_+\backslash\{0\}$ by the relevance,
it suffices to see that $\rho(-\olm) \leq 0$ for any $\olm\in\olM$.
If there exists an $\olm\in\olM$ with $\rho(-\olm)>0$,
then we can find a $Q\in\calQ$ such that
$\bbE_Q[\olm]>\rho^*(Q)\geq(\rho^0)^*(Q)=\sup_{m \in M}\bbE_Q[m]$
by Proposition \ref{prop2-1}.
This is a contradiction.

\noindent
2$\Rightarrow$4:
Since condition 2 implies condition 1,
$\wh{\rho^0}$ is a GDV by Theorem \ref{thm1}.
Next, we show the relevance of $\wh{\rho^0}$.
For any $z\in L_+\backslash\{0\}$, $z\wedge1$ belongs to $B_\delta$
for some $\delta\in(0,1]$, where $B_\delta$ is defined in (\ref{B-delta}).
Since $B_\delta\cap\olM=\emptyset$,
Proposition \ref{prop-sep} implies the existence of $Q\in\calQ$ satisfying
$\sup_{\olm\in\olM}\bbE_Q[\olm]<\inf_{x\in B_\delta}\bbE_Q[x]$.
Then, we have $0\leq\sup_{\olm\in\olM}\bbE_Q[\olm]
<\bbE_Q[z\wedge1]\leq\bbE_Q[z]$.
Consequently, $\wh{\rho^0}$ is relevant.

\noindent
1$^\prime\Rightarrow$4:
Theorem \ref{thm1} and Lemma \ref{lem0-2-4} imply
that $\wh{\rho^0}$ is a relevant GDV.
\fin

\begin{rem}
We can regard Theorem \ref{thm3} as a generalization of
Corollary 9.32 in \cite{FS11}.
\end{rem}

In order to complete Theorem \ref{thm3},
we illustrate counterexamples for the implications which are not
shown.

\begin{ex}[Counterexample for 1$\Rightarrow$2]
\label{counter-ex1}
Setting $\Omega=\{\omega_k|k\in\bbN\}$ and
$\bbP(\{\omega_k\})=2^{-k}$ for each $k\in\bbN$,
we define random variables $S_k$, $k\in\bbN$ as
\[
S_k(\omega)=\l\{
            \begin{array}{ll}
               1, & \mbox{ if }\omega=\omega_k, \\
               0, & \mbox{ if }\omega\neq\omega_k, \\
            \end{array}\r.
\]
and $M=\co\{S_1,S_2,\dots\}-L_+$.
Remark that any element $m\in\co\{S_1,S_2,\dots\}$ is expressed as
$m=\sum_{k=1}^\infty\lambda_kS_k$,
where the sequence $\{\lambda_k\}_{k\in\bbN}$ satisfies
$\lambda_k\geq0$ for any $k\geq1$, $\sum_{k=1}^\infty\lambda_k=1$ and
$\lambda_k=0$ except for finitely many $k$s.
This model then does not satisfy condition 2.

Next, we make sure of condition 1.
To this end, we define for each $n\in\bbN$,
\[
Q_n(\{\omega_k\}):=\l\{
            \begin{array}{ll}
               \frac{1}{n},          & \mbox{ if }k\leq n-1, \\
               \frac{2^{n-k-1}}{n},  & \mbox{ otherwise}. \\
            \end{array}\r.
\]
We can see that each $Q_n$ is a probability measure equivalent to $\bbP$; and
$\frac{\d Q_n}{d\bbP}\leq\frac{2^{n-1}}{n}$.
For $m\in\co\{S_1,S_2,\dots\}$ with $m=\sum_{k=1}^\infty\lambda_kS_k$, we have
\[
\bbE_{Q_n}[m]=\sum_{k=1}^\infty\lambda_kQ_n(\{\omega_k\})
\leq\sum_{k=1}^\infty\frac{\lambda_k}{n}=\frac{1}{n}.
\]
Thus, we obtain, for any $n\in\bbN$,
\[
(\rho^0)^*(Q_n)=    \sup_{m\in M}\bbE_{Q_n}[m]
               =    \sup_{m\in\co\{S_1,S_2,\dots\}}\bbE_{Q_n}[m]
               \leq \frac{1}{n},
\]
from which condition 1 follows.
\end{ex}

\begin{ex}[Counterexample for 4$\Rightarrow$1$^\prime$]
\label{counter-ex2}
We take $\Omega=\{\omega_k|k\in\bbZ\}$ and
a probability measure $\bbP$ with $\bbP(\{\omega_k\})>0$ for each $k\in\bbZ$.
Further, we define random variables $S_k$, $k\in\bbZ$ as
\[
S_k(\omega)=\l\{
            \begin{array}{ll}
                1, & \mbox{ if }\omega=\omega_k, \\
               -1, & \mbox{ if }\omega=\omega_{k-1}, \\
                0, & \mbox{ otherwise}, \\
            \end{array}\r.
\]
and $M=\co\{S_k|k\in\bbZ\}-L_+$.

Now, we see that this model satisfies condition 4.
We define, for each $i\in\bbZ$ and $j\in\bbN$,
\[
Q^i_j(\{\omega_k\}):=\l(\frac{1}{j}-\frac{|k-i|}{j^2}\r)\vee0.
\]
We can see that each $Q^i_j$ is a probability measure
with bounded density $\frac{\d Q^i_j}{\d\bbP}$.
For any $m\in M$ with representation $m=\sum_{k=1}^\infty\lambda_kS_k$,
we have
\[
\bbE_{Q^i_j}[m]=\sum_{k=1}^\infty\lambda_k\bbE_{Q^i_j}[S_k]
=\sum_{k=1}^\infty\lambda_k\l\{Q^i_j(\{\omega_k\})-Q^i_j(\{\omega_{k-1}\})\r\}
\leq\frac{1}{j^2}.
\]
We have then $(\rho^0)^*(Q^i_j)=\frac{1}{j^2}$, that is, $Q^i_j\in\calQ$.
Thus, $\calQ\neq\emptyset$ and $\inf_{Q\in\calQ}(\rho^0)^*(Q)=0$,
which ensure that $\wh{\rho^0}$ is a GDV.
On the other hand, taking a $z\in L_+\backslash\{0\}$ arbitrarily,
we can find an $\ve>0$ and an $i\in\bbZ$ satisfying $z\geq\ve1_{\{\omega_i\}}$.
Hence, we have $\bbE_{Q^i_j}[z]\geq\ve Q^i_j(\{\omega_i\})=\frac{\ve}{j}$
for any $j\in\bbN$.
For a sufficient large $j$, we have $\bbE_{Q^i_j}[z]>(\rho^0)^*(Q^i_j)$,
that is, $\wh{\rho^0}(-z)>0$.

Next, we see that condition 1$^\prime$ does not hold.
Since condition 4 holds, so does condition 1 by Theorem \ref{thm3},
that is, $\calQ^e$ is nonempty.
For any $Q\in\calQ^e$, there exists a $k_Q\in\bbZ$ such that
$Q(\{\omega_{k_Q}\})-Q(\{\omega_{k_Q-1}\})>0$.
Indeed, since $K_0:=\{k\in\bbZ|Q(\{\omega_k\})\geq Q(\{\omega_0\})\}$
is finite, we can take $k_Q=\min K_0$.
Hence, we have, for any $Q\in\calQ^e$,
\[
(\rho^0)^*(Q)=    \sup_{m\in M}\bbE_Q[m]
             \geq \bbE_Q[S_{k_Q}]
             =    Q(\{\omega_{k_Q}\})-Q(\{\omega_{k_Q-1}\})
             >    0,
\]
which denies condition 1$^\prime$.
\end{ex}

Here we give an equivalent condition to condition 1$^\prime$
in Theorem \ref{thm3}.

\begin{prop}
There exists a relevant coherent GDV if and only if
$\calQ^e\cap\calQ_0$ is nonempty.
\end{prop}

\proof
``if" part: \ 
This is shown by a similar way with Proposition \ref{prop-coh}.
Taking a $Q\in\calQ^e\cap\calQ_0$,
we define $\rho_Q(x):=\bbE_Q[-x]$ for any $x\in L$.
Then, $\rho_Q$ is a coherent GDV which is relevant.

``only if" part: \ 
This follows from the Halmos-Savage theorem
(see e.g. Theorem 25 in \cite{Del}).
Now, we give just a sketch of proof.

Let $\rho$ be a relevant coherent GDV.
Denoting $A:=\{x\in L|\rho(-x)\leq0\}$, we have
$\sup_{x\in A}\bbE_Q[x]=\rho^*(Q)$ for any $Q\in\calQ$.
Now, we consider $B_\delta$ defined in (\ref{B-delta}).
Since $A\cap B_\delta=\emptyset$ for any $\delta\in(0,1]$,
the same separating argument as Proposition \ref{prop-sep} implies that,
for any $\delta\in(0,1]$, there exists a $Q\in\calQ$ such that
\[
\rho^*(Q)=\sup_{x\in A}\bbE_Q[x]<\inf_{z\in B_\delta}\bbE_Q[z],
\]
since $A$ is $\sigma(L,L^\Psi)$-closed.
Now, for each $\delta\in(0,1]$, we denote
\[
\calQ_\delta:=\{Q\in\calQ|\inf_{z\in B_\delta}\bbE_Q[z]>\rho^*(Q)\}.
\]
Remark that $\calQ_\delta$ is stable for countable unions.
Then, we can find a $Q_\delta\in\calQ_\delta$ satisfying
$\bbP(\{\frac{\d Q_\delta}{\d\bbP}>0\})
=\max_{Q\in\calQ_\delta}\bbP(\{\frac{\d Q}{\d\bbP}>0\})$.
In addition, we can see that $\bbP(\{\frac{\d Q_\delta}{\d\bbP}>0\})=1$
by contradiction.
Since $\delta\in B_\delta$, we have $\rho^*(Q_\delta)<\delta$,
from which $\rho^*(Q_\delta)=0$ holds,
since $\rho^*$ takes the values $0$ and $\infty$ only.
Hence, $(\rho^0)^*(Q_\delta)=0$ by Proposition \ref{prop2-1},
that is, $Q_\delta\in\calQ^e\cap\calQ_0$.
\fin

\subsection{An extension theorem}
We assume that any $x\in L$ is priced at $\rho(-x)$,
where $\rho$ is a GDV.
Then $x-\rho(-x)$ is a $0$-attainable claim.
Now, we extend our market by adding all these claims to $M$.
More precisely, the set of $0$-attainable claims for the extended market
is represented as
\begin{eqnarray*}
M^\rho
&:=& \{x-\rho(-x)|x\in L, \rho(-x)<\infty\}-L_+ \\
&=&  \{x\in L|\rho(-x)=0\}-L_+
     =  \{x\in L|\rho(-x)\leq0\}.
\end{eqnarray*}
Remark that $M^\rho$ is a convex set including $M$.
Since $M^\rho$ is closed in $\sigma(L,L^\Psi)$ by Theorem \ref{BF},
the NFL for the extended market is equivalent to $M^\rho\cap L_+=\{0\}$,
which is the no-arbitrage condition.
We have the following theorem:

\begin{thm}
\label{thm4}
Let $\rho$ be a GDV.
The following are equivalent:
\begin{enumerate}
\item $\rho$ is relevant.
\item $-\wh{\rho^0}(x-z)<\rho(-x)$ for any $x \in L$
      and $z\in L_+\backslash \{0\}$.
\item $-\rho^0(x-z)<\rho(-x)$ for any $x\in L$
      and $z\in L_+\backslash\{0\}$.
\item $M^\rho\cap L_+=\{0\}$.
\end{enumerate}
\end{thm}

\proof
The implications 2$\Rightarrow$3$\Rightarrow$1$\Leftrightarrow$4
are shown by the same way as Theorem 4.3 in \cite{AF}.
Then we have only to see the implication 1$\Rightarrow$2.
For any $z \in L_+\backslash\{0\}$,
there exists $Q_z\in\calQ$ such that $\bbE_{Q_z}[z/2]>\rho^*(Q_z)$ 
by the relevance of $\rho$.
Thus, Proposition \ref{prop2-1} implies that
$\bbE_{Q_z}[z]>2\rho^*(Q_z)\geq\rho^*(Q_z)+(\rho^0)^*(Q_z)$.
Therefore, for any $x\in L$, we have
\begin{eqnarray*}
-\wh{\rho^0}(x-z)
&=& \inf_{Q \in \calQ}\l\{\bbE_Q[x-z]+(\rho^0)^*(Q)\r\}
    \leq \bbE_{Q_z}[x-z]+(\rho^0)^*(Q_z) \\
&<& \bbE_{Q_z}[x]-\rho^*(Q_z)
    \leq \sup_{Q\in\calQ}\l\{\bbE_Q[x]-\rho^*(Q)\r\}
    =    \rho(-x).
\end{eqnarray*}
\fin

\setcounter{equation}{0}
\section{Conclusions}
We study properties of good deal bounds for incomplete markets with
convex constraints.
In Section 3, we study properties of superhedging cost $\rho^0$ and
its largest minorant with the Fatou property $\wh{\rho^0}$.
Next, we see that the existence of a GDV is equivalent to
``$\calQ\neq\emptyset$ and $\inf_{Q\in\calQ}(\rho^0)^*(Q)=0$"
in Theorem \ref{thm1};
and enumerate equivalent conditions for a given $\rho\in\calR$ to be a GDV
in Proposition \ref{prop2-1}.
Moreover, we introduce an example of a GDV which is not a risk indifference
price; and look into relationship between GDVs and risk indifference prices.
Furthermore, we prove an FTAP under convex constraints in Theorem \ref{thm3}.
Among others, the equivalence between the NFL and the existence of a relevant
GDV is proved.
Moreover, we illustrate counterexamples to see that
neither $\calQ^e\cap\calQ_0\neq\emptyset$ nor
$\inf_{Q\in\calQ^e}(\rho^0)^*(Q)=0$ is equivalent to the NFL.



\begin{thebibliography}{}
\bibitem{A11}
        {\sc Arai, T.} (2011)
        \emph{Good deal bounds induced by shortfall risk},
        SIAM J. Financial Math., {\bf 2}, 1-21.
\bibitem{AF}
        {\sc Arai, T. and Fukasawa, M.} (2014)
        \emph{Convex risk measures for good deal bounds},
        Mathematical Finance, {\bf 24}, 464-484.
\bibitem{BF09}
        {\sc Biagini, S. and Frittelli, M.} (2009)
        \emph{On the extension of the Namioka-Klee theorem and
        on the Fatou property for risk measures},
        in Optimality and risk: modern trends in mathematical finance,
        The Kabanov Festschrift, Delbaen, Rasonyi, Stricker eds., 1-28,
        Springer.
\bibitem{BFG}
        {\sc Biagini, S., Frittelli, M. and Grasselli, M.} (2011)
        \emph{Indifference price with general semimartingales},
        Mathematical Finance, {\bf 21}, 423-446.\bibitem{BN09}
        {\sc Bion-Nadal, J.} (2009)
        \emph{Bid-Ask dynamic pricing in financial markets with
        transaction costs and liquidity risk},
        Journal of Mathematical Economics, {\bf 45}, 738-750.
\bibitem{BNDN}
        {\sc Bion-Nadal, J. and Di Nunno, G.} (2013)
        \emph{Dynamic no-good-deal pricing measures and extension theorems
        for linear operators on $L^\infty$},
        to appear in Finance and Stochastics.
\bibitem{CPT01}
        {\sc Carassus, L., Pham, H. and Touzi, N.} (2001)
        \emph{No arbitrage in discrete time under portfolio constraints},
        Mathematical Finance, {\bf 11}, 315-329.
\bibitem{CL09}
        {\sc Cheridito, P. and Li, T.} (2009)
        \emph{Risk measures on Orlicz hearts},
        Mathematical Finance, {\bf 19}, 189-214.
\bibitem{CS00}
        {\sc Cochrane, J.H. and Sa\'a-Requejo, J.} (2000)
        \emph{Beyond arbitrage: good-deal asset price bounds
        in incomplete markets},
        Journal of Political Economy, {\bf 108}, 79-119.
\bibitem{Cuoco}
        {\sc Cuoco, D.} (1997)
        \emph{Optimal consumption and equilibrium prices with portfolio
        constraints and stochastic income},
        Journal of Economic Theory, {\bf 72}, 33-73.
\bibitem{CK92}
        {\sc Cvitani\'c, J. and Karatzas, I.} (1992)
        \emph{Convex duality in constrained portfolio optimization},
        Annals of Applied Probability, {\bf 2}, 767-818.
\bibitem{CK93}
        {\sc Cvitani\'c, J. and Karatzas, I.} (1993)
        \emph{Hedging contingent claims with constrained portfolios},
        Annals of Applied Probability, {\bf 3}, 652-681.
\bibitem{Del} 
        {\sc Delbaen, F. } (2012)
        \emph{Monetary Utility Functions},
        Osaka University CSFI Lecture Notes Series, Osaka University Press.
\bibitem{DS06}
        {\sc Delbaen, F. and Schachermayer, W.} (2006)
        \emph{The mathematics of arbitrage},
        Springer Finance. Springer-Verlag, Berlin.
\bibitem{ES92}
        {\sc Edgar, G.A. and Sucheston, L.} (1992)
        \emph{Stopping times and directed processes},
        Cambridge University Press.
\bibitem{EST04}
        {\sc Evstigneev, I.V., Sch\"urger, K. and Taksar, M.I.}(2004)
        \emph{On the fundamental theorem of asset pricing: 
        random constraints and Bang-Bang no-arbitrage criteria},
        Mathematical Finance, {\bf 14}, 201-221.
\bibitem{FK97}
        {\sc F\"ollmer, H. and Kramkov, D.} (1997)
        \emph{Optional decompositions under constraints},
        Probability Theory and Related Fields, {\bf 109}, 1-25.
\bibitem{FS11}
        {\sc F\"ollmer, H. and Schied, A.} (2011)
        \emph{Stochastic Finance 3rd edition},
        De Gruyter, Berlin.
\bibitem{JK01}
        {\sc Jaschke, S. and K\"uchler, U.} (2001)
        \emph{Coherent risk measures and good-deal bounds},
        Finance and Stochastics, {\bf 5}, 181-200.
\bibitem{KK96}
        {\sc Karatzas, I. and Kou, S.} (1996)
        \emph{On the pricing of contingent claims under constraints},
        Annals of Applied Probability, {\bf 6}, 321-369.
\bibitem{K12}
        {\sc Kardaras, C.} (2012)
        \emph{Market viability via absence of arbitrage of the first kind},
        Finance and Stochastics, {\bf 16}, 651-667.
\bibitem{KS05}
        {\sc Kl\"oppel, S. and Schweizer, M.} (2005)
        \emph{Dynamic indifference valuation via convex risk measures},
        NCCR FINRISK working paper No.209, ETH Z\"urich..
\bibitem{K81}
        {\sc Kreps, D.M.} (1981)
        \emph{Arbitrage and equilibrium in economies with infinitely many
        commodities},
        Journal of Mathematical Economics, {\bf 8}, 15-35.
\bibitem{LZ13}
        {\sc Larsen, K. and \v{Z}itkovi\'{c}, G.} (2013)
        \emph{On utility maximization under convex portfolio constraints},
        Annals of Applied Probability, {\bf 23}, 665-692.
\bibitem{MC}
        {\sc Madan, D. and Cherny, A.} (2010)
        \emph{Markets as a counterparty: an introduction to conic finance},
        International Journal of Theoretical and Applied Finance,
        {\bf 13}, 1149-1177.
\bibitem{P1}
        {\sc Pennanen, T.} (2011)
        \emph{Arbitrage and deflators in illiquid markets},
        Finance and Stochastics, {\bf 15}, 57-83.
\bibitem{P2}
        {\sc Pennanen, T.} (2011)
        \emph{Superhedging in illiquid markets},
        Mathematical Finance, {\bf 21}, 519-540.
\bibitem{PP}
        {\sc Pennanen, T. and Penner, I.} (2010)
        \emph{Hedging of claims with physical delivery
        under convex transaction costs},
        SIAM Journal on Financial Mathematics, {\bf 1}, 158-178.
\bibitem{RR}
        {\sc Rao, M.M. and Ren, Z.D.} (1991)
        \emph{Theory of Orlicz spaces},
        Marcel Dekker, Inc., New York.
\bibitem{R05}
        {\sc Rokhlin, D.B.} (2005)
        \emph{An extended version of the Dalang-Morton-Willinger theorem
        under portfolio constraints},
        Theory of Probability \& Its Applications, {\bf 49}, 429-443.
\bibitem{R11}
        {\sc Roux, A.} (2011)
        \emph{The fundamental theorem of asset pricing in the presence of
        bid-ask and interest rate spreads},
        Journal of Mathematical Economics, {\bf 47}, 159-163.
\bibitem{Simons}
        {Simons, S.} (1999)
        \emph{Minimax and Monotonicity},
        Lecture Notes in Mathematics, {\bf 1693}, Springer.
\bibitem{S04}
        {\sc Staum, J.} (2004)
        \emph{Fundamental theorems of asset pricing for good deal bounds},
        Mathematical Finance, {\bf 14}, 141-161. 
\bibitem{X06}
        {\sc Xu, M.} (2006)
        \emph{Risk measure pricing and hedging in incomplete markets},
        Annals of Finance, {\bf 2}, 51-71.
\end{thebibliography}
\end{document}